\documentclass[conference]{IEEEtran}
\IEEEoverridecommandlockouts
\usepackage{cite,url,hyperref,wrapfig,subcaption}
\usepackage{amsmath,amssymb,amsfonts}
\usepackage{algorithmic}
\usepackage{tabularray}

\usepackage{graphicx}
\usepackage{textcomp}
\usepackage[normalem]{ulem} %gulabi
\usepackage{xcolor,float}

\usepackage[nolist,nohyperlinks]{acronym}

\newcommand{\gm}[1]{{\color{black}{#1}}}
\newcommand{\ar}[1]{{\color{black}{#1}}}
\newcommand\blfootnote[1]{%
  \begingroup
  \renewcommand\thefootnote{}\footnote{#1}%
  \addtocounter{footnote}{-1}%
  \endgroup
}
\def\BibTeX{{\rm B\kern-.05em{\sc i\kern-.025em b}\kern-.08em
    T\kern-.1667em\lower.7ex\hbox{E}\kern-.125emX}}
    
\begin{document}

% \title{A Study on Radio Resource Allocation for Different Vehicle Platoon Control Algorithms in 5G eV2X
% % {\footnotesize \textsuperscript{*}Note: Sub-titles are not captured for https://ieeexplore.ieee.org  and
% % should not be used}
% % \thanks{Identify applicable funding agency here. If none, delete this.}
% }
\title{Performance Analysis of Resource Allocation Algorithms for Vehicle Platoons over 5G eV2X Communication 
% {\footnotesize \textsuperscript{*}Note: Sub-titles are not captured for https://ieeexplore.ieee.org  and
}

% \author{\IEEEauthorblockN{1\textsuperscript{st}Gulabi Mandal}
% \IEEEauthorblockA{\textit{Department of Computer Science and Engineering} \\
% \textit{Indian Institute of Technology}\\
% Kharagpur, India\\
% gulabi007@kgpian.iitkgp.ac.in}
% \and
% \IEEEauthorblockN{2\textsuperscript{nd}Anik Roy}
% \IEEEauthorblockA{\textit{Advanced Technology Development Center} \\
% \textit{Indian Institute of Technology}\\
% Kharagpur, India\\
% anikroy@kgpian.iitkgp.ac.in}
% \and
% \IEEEauthorblockN{3\textsuperscript{rd}Basabdatta Palit}
% \IEEEauthorblockA{\textit{Electronics and Communication Engineering} \\
% \textit{National Institute of Technology}\\
% Rourkela, India \\
% palitb@nitrkl.ac.in}
% }
 \author{\IEEEauthorblockN{Gulabi Mandal$\textbf{}^1$, Anik Roy$\textbf{}^2$, Basabdatta Palit$\textbf{}^3$}
\IEEEauthorblockA{$\textbf{}^1$Department of Computer Science and Engineering,  
Indian Institute of Technology, Kharagpur, India\\$\textbf{}^2$Advanced Technology Development Center, Indian Institute of Technology, Kharagpur, India\\
$\textbf{}^3${Department of Electronics and Communication Engineering,  National Institute of Technology, Rourkela} }
}

\maketitle

\begin{abstract}
Vehicle platooning is a cooperative driving technology that can be supported by 5G enhanced Vehicle-to-Everything (eV2X) communication to improve road safety, traffic efficiency, and reduce fuel consumption.  eV2X communication among the platoon vehicles involves the periodic exchange of Cooperative Awareness Messages (CAMs) containing vehicle information under strict latency and reliability requirements. These requirements can be maintained by administering the assignment of resources, in terms of time slots and frequency bands, for CAM exchanges in a platoon, with the help of a resource allocation mechanism. State-of-the-art on control and communication design for vehicle platoons either consider a simplified platoon model with a detailed communication architecture or consider a simplified communication delay model with a detailed platoon control system. 
Departing from existing works, we have developed a comprehensive vehicle platoon communication and control framework using $\mathtt{OMNET++}$, the benchmarking network simulation tool. We have carried out an inclusive and comparative study of three different platoon Information Flow Topologies (IFTs), namely Car-to-Server, Multi-Hop, and One-Hop over 5G using the Predecessor-leader following platoon control law to arrive at the best-suited IFT for platooning. Secondly, for the best-suited 5G eV2X platooning IFT selected, we have analyzed the performance of three different resource allocation algorithms, namely Maximum of Carrier to Interference Ratio (MaxC/I), Proportional Fair (PF), and Deficit Round Robin (DRR). Exhaustive system-level simulations show that the One-Hop information flow strategy along with the MaxC/I resource allocation yields the best Quality of Service (QoS) performance, in terms of latency, reliability, Age of Information (AoI), and throughput. \blfootnote{Authors Gulabi Mandal and Anik Roy are equal contributors. This work was carried out when they were MTech students at IIEST, Shibpur, and Dr. Basabdatta Palit was a faculty at IIEST Shibpur.}
\end{abstract}

\acresetall
% \begin{figure*}[hbt!]
%   \includegraphics[width=\textwidth,height=4cm]{Images/Architectures.png}
%    \caption{CAM exchange within a platoon in (a) Car-to-Server (b) Multi-Hop (c) One-Hop mode.}
%    \label{Fig:PltnArchitectures}
% \end{figure*}

\section{Introduction}
\label{sec:Intro}
The unprecedented growth in the global density of automobiles in the last decade has triggered an increase in the carbon footprint while compromising on-road safety.  For example, in India, while the private ownership of four-wheeler vehicles increased by 11\% from 2011 to 2015, the contribution to carbon emission only from passenger vehicles also became nearly 45\%, over the same period. Besides, the number of accidents also increased to as high as 449,002 in 2019 in India leading to 151,113 deaths~\cite{RoadAccidentStat2019}. 
The panacea for all such problems can be the \acp{CAV}~\cite{Loke2019}. % CAV cooperate amongst themselves to improve on-road traffic efficiency and safety through coordinated driving, route selection, or car parking. These are also automated in terms of route correction, adjusting to the road demographics as well as to the dynamics of the neighbouring vehicles. 
A potential use case of \ac{CAV} is grouping vehicles into a platoon, which is projected to increase the road traffic capacity through coordinated driving while reducing the carbon footprint.   

A vehicle platoon is essentially a train of vehicles with a common interest, moving in unison with a small safety gap in between. The members in a vehicle platoon move in a coordinated fashion such that they can accelerate or decelerate coherently, resulting in considerable savings in fuel consumption due to the reduced aerodynamic drag. Coordinated driving requires that the \textit{string stability} of the vehicle platoons be guaranteed in order to improve on-road traffic efficiency and road safety \cite{Feng2019, TStrurm2020}. This is effectuated through \ac{CACC}, a vehicular control strategy. This in turn requires a periodic, timely, and reliable exchange of \ac{CAM} \cite{Nardini2018} which contain essential information, such as current positions of the vehicles, changes in speed and acceleration, etc. 
 \\
\indent The timely delivery of \acp{CAM}  between the \ac{PL} and \acp{PM} depends on the underlying \ac{V2V} communication links, which are essentially lossy in nature, primarily due to multipath fading and Doppler spread\cite{MoradiPari2023, 3GPP_MultiPath_fading_2021}. A popular standard for \ac{V2V} communication is the IEEE 802.11p \ac{DSRC}~\cite{Kenney2011}. However, it fails to deliver the stringent latency requirements associated with vehicle platoons in 4G and beyond networks~\cite{Gorospe2024}. So, \ac{3GPP} prescribed the \ac{cV2X} communication standard in LTE, which executes the \ac{V2V} communication over the existing \ac{4G} cellular infrastructure~\cite{3GPPPathloss_R15_2019, Nardini2020}. \ac{cV2X} can support longer communication distances between heterogeneous network nodes, such as \ac{V2V}, \ac{V2I}, and \ac{V2P}~\cite{Bey2018,Boubakri2020}. With the requirement for ultra-reliable low latency communication among vehicles in 5G, \ac{3GPP} has extended \ac{cV2X} to the \ac{eV2X} standard in Release-16~\cite{3GPP2019Rel16}. Nonetheless, with 5G promising to operate in the higher frequency range, the signal quality fluctuations in the \ac{V2V} links are expected to increase further due to their increased susceptibility to Doppler spread and multipath fading. One of the methods to counteract this high-frequency channel impairments and improve the reliability of the communication links is to design efficient radio resource allocation methods.  

\indent As in \ac{cV2X}, Release 16 also recommends \ac{OFDMA} for \ac{eV2X}, albeit with a multi-numerology frame structure. In multi-numerology \ac{OFDMA}, the bandwidth is divided into orthogonal \acp{BWP}, and in each \ac{BWP} the time-frequency \acp{RB} have different bandwidths and time duration to cater to different applications. A judicious allocation of these \acp{RB} to serve the \acp{CAM} in a \ac{TTI} can ensure reliable and timely delivery of these messages.   %The number of \acp{RB} needed by a user depends on their underlying channel conditions, packet size, and \ac{QoS} requirements - this method is called link adaptation. 
\ac{3GPP} does not specify any \ac{RB} allocation algorithm for any of the standards. So, it is important to design such algorithms for efficient communication in a vehicle platoon. %The ability of these algorithms to dynamically allocate \acp{RB} to the different users depending on their channel conditions and required QoS helps in the timely delivery of \acp{CAM}, which in turn improves the stability of the platoons. 
\subsection{Related Works} The prior art on \ac{V2V} communication over vehicle platoons can be grouped into two categories. The first category focuses on the performance analysis of \ac{cV2X}-based platoon communication, with studies focusing on transmission latency~\cite{Nardini2020,He2023,GNardini2016,Nardini2016}, and communication reliability~\cite{Zeng2019,Wen2020,Lakshmanan2021,Liu2024}. The second category focuses on resource allocation algorithms for platoons, which can be further divided into two subgroups, (i) performance evaluation of existing resource allocation algorithms like \ac{MaxC/I} and \ac{DRR}~\cite{Nardini2018,Palash2024}, and (ii) the design of resource allocation strategies aimed at reducing latency~\cite{Peng2017,Wang2018,Wang2019,Mei2018,Zhang2022,Cao2023,Chai2024,Lei2024}. However, these existing studies~\cite{Nardini2020,He2023,GNardini2016,Nardini2016,Zeng2019,Wen2020,Lakshmanan2021} do not take into consideration different types of communication (information) flows in platoons and their impact on the latency and reliability performance. Furthermore, the resource allocation works~\cite{Peng2017,Wang2018,Wang2019,Mei2018,Zhang2022,Cao2023,Chai2024,Lei2024} solely focus on improving the latency performance while considering a fixed communication flow. The communication flows as highlighted in~\cite{GNardini2016,Nardini2016}, and~\cite{Naik2019} can substantially impact the latency and reliability performance of platoons, which in turn depends on the resource allocation algorithms under use. Clearly, a detailed performance analysis of resource allocation algorithms for platoon configuration, particularly those that achieve desirable latency profiles in 5G \ac{eV2X}-based vehicular communication, remains unexplored in the existing literature.

\begin{figure*}[htbt!]
    \centering
    \begin{subfigure}[b]{0.32\textwidth}
    \centering
    \includegraphics[width=\linewidth,clip]{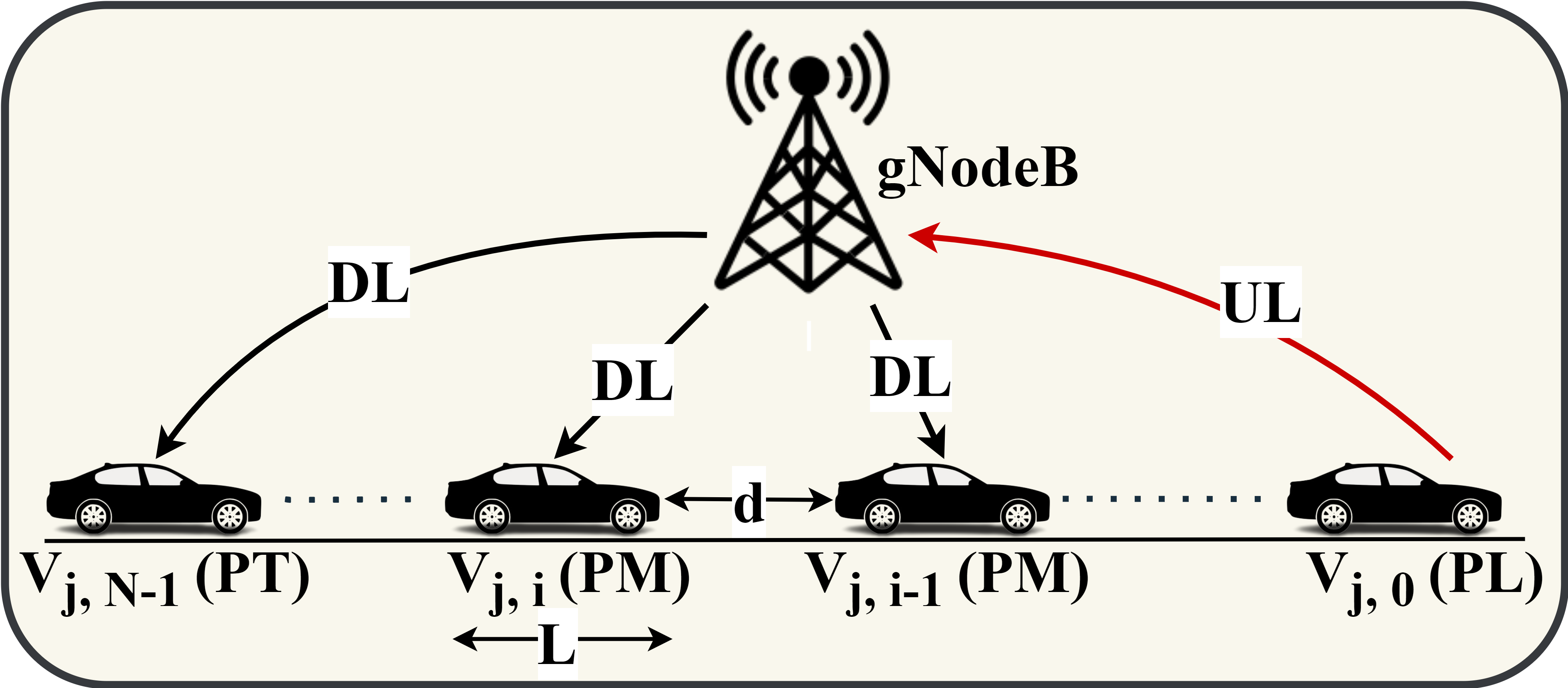}
    \caption{}
    \label{fig:Car2Server}
    \end{subfigure}
    \hfill % Add horizontal space between figures
    \begin{subfigure}[b]{0.32\textwidth}
    \centering
    \includegraphics[width=\linewidth,clip]{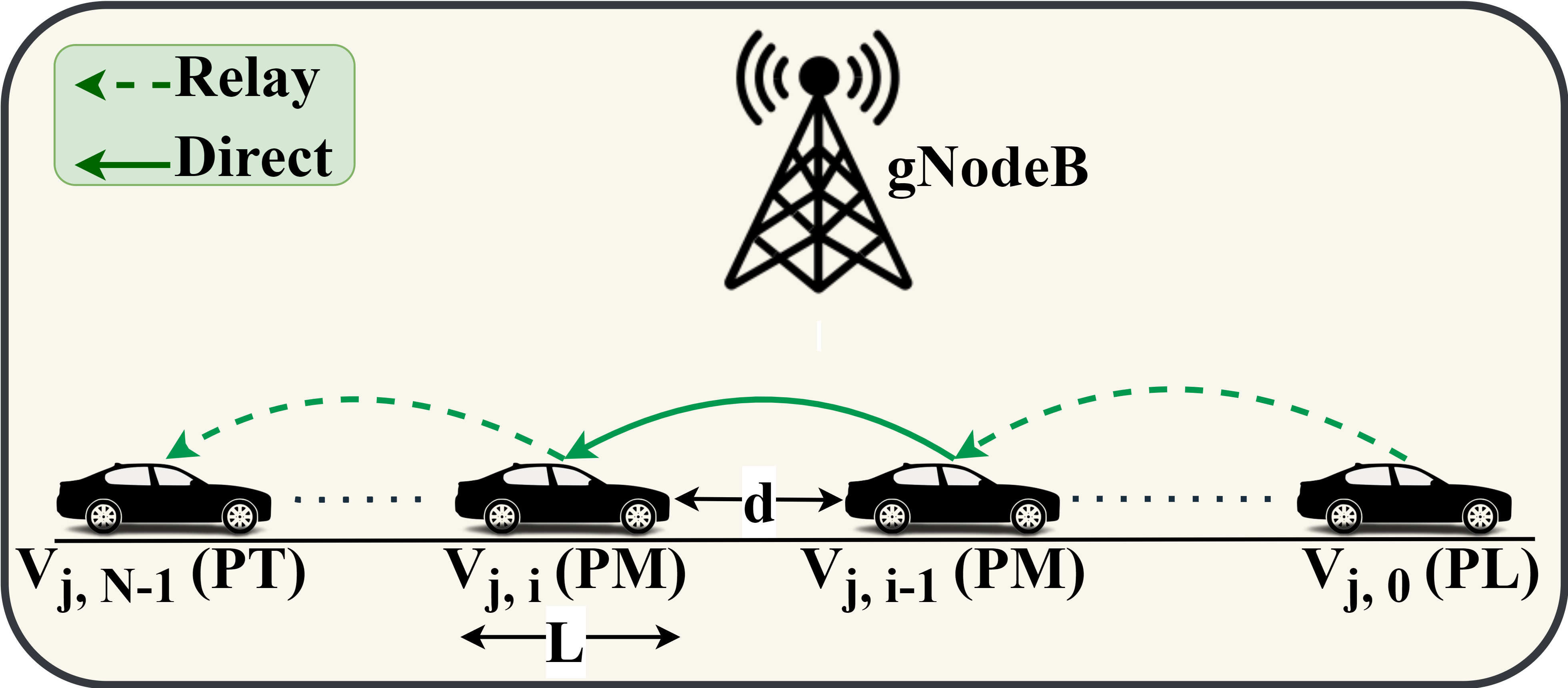}
    \caption{}
    \label{fig:MultiHop}
    \end{subfigure}
    \hfill % Add horizontal space between figures
    \begin{subfigure}[b]{0.32\textwidth}
    \centering
    \includegraphics[width=\linewidth,clip]{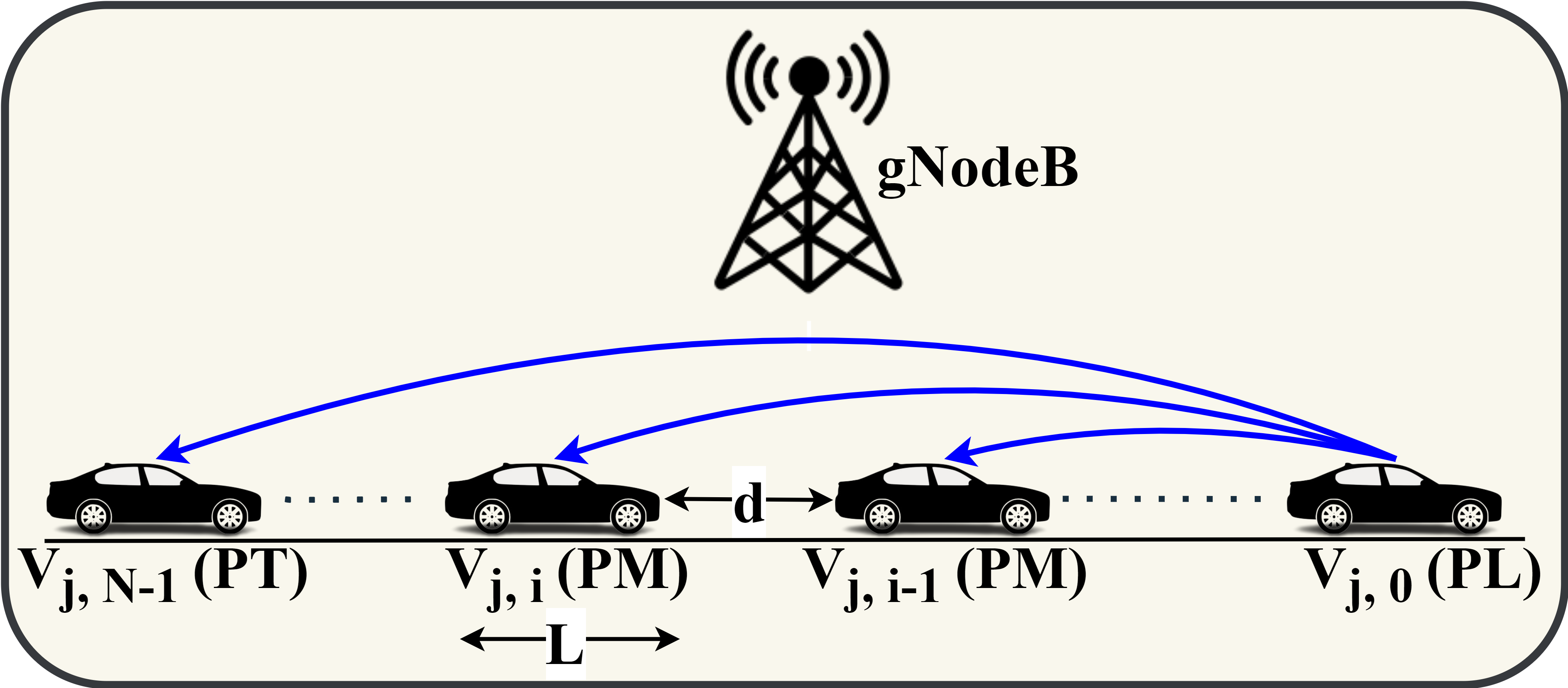}
    \caption{}
    \label{fig:OneHop}
    \end{subfigure}
    % \vspace{-0.4cm}
    \caption{Platoon \acp{IFT} for CAM: (a) Car-to-Server, (b) Multi-Hop, and (c) One-Hop.}
    \label{Fig:PltnArchitectures}
    % \vspace{-0.7cm}
\end{figure*}

\subsection{Contributions} In this work, we have undertaken a comprehensive study on the different resource allocation algorithms that can be used to improve the performance of vehicular platoons in 5G. Existing works on platoon management either consider a detailed control law with a simplified communication model~\cite{Zhou2023,Guo2023,Shen2023} or consider an abstraction of a platoon with a well-developed communication protocol~\cite{Huang2023,Chai2023,Chai2024}. In contrast, in this work, we have used a well-defined control flow topology in addition with the complete protocol stack of 5G using a combindation of the $\mathtt{INET}$, $\mathtt{Veins}$, $\mathtt{PLEXE}$, and $\mathtt{Simu5G}$ libraries of the benchamrking $\mathtt{OMNET++}$ network simulator. We have used the default \ac{PLF} control strategy of $\mathtt{PLEXE}$~\cite{Segata2014}, where the control law of the \ac{PM} updates at regular intervals based on the \ac{CAM} received from its predecessor and \ac{PL}. To ensure timely reception of the \ac{CAM}, we have used multi-numerology \ac{OFDMA}-based \ac{V2V} communication over 5G, which has been implemented using $\mathtt{Simu5G}$~\cite{Nardini2020}. The information flows from the \ac{PL} and the predecessor to the \acp{PM} by using one of the three \ac{IFT} - \ac{V2V} communication via infrastructure (Car-to-Server), direct \ac{V2V} communication (One-Hop), and relayed \ac{V2V} communication (Multi-Hop).
% In contrast, in this work, we have considered a detailed vehicle platoon control and \ar{\ac{IFT}} built using the $\mathtt{INETs}$, $\mathtt{Simu5G}$, $\mathtt{Veins}$ and $\mathtt{PLEXE}$ libraries of the $\mathtt{OMNeT++}$ network simulator. \ar{We have considered the \ac{PLF}-based \ac{CACC} for the platoon control model. Each \ac{PM} receives its \ac{PL} information using three \acp{IFT} that we have considered in this work - V2V communication via infrastructure (Car-to-Server), direct V2V communication (One-Hop), and relayed V2V communication (Multi-Hop)}.
% We have considered three \acp{IFT}, i.e., three platoon communication architectures - V2V communication via infrastructure (Car-to-Server), direct V2V communication from \ac{PL} to all \acp{PM} (One-Hop), and relayed V2V communication from \ac{PL} to the \acp{PM} (Multi-Hop). 
Then, we have investigated how resource allocation strategies, such as \ac{MaxC/I}, \ac{PF}, and \ac{DRR}~\cite{Zain2015} affect the information topologies. 

As discussed earlier, platoons can improve road traffic capacity. Although a longer platoon can streamline road traffic better, the length can negatively impact stability. So, in this work, we have investigated how different resource allocation algorithms affect the reliability of \ac{CAM} transmissions in platoons with longer lengths. The reliability is quantified in this work in terms of metrics like average \ac{E2E} delay, average \ac{AoI}, average throughput, and reception probability~\cite{Rahim2019}. 
% In addition, we have also studied how the different resource allocation algorithms can support multiple platoons in the same bandwidth. 
Thus, the contributions of our work are as follows.
\begin{itemize}
    \item We have designed an end-to-end system model for a vehicle platoon framework considering different \acp{IFT} like Car-to-Server, One-Hop, and Multi-Hop, for \ac{V2V} communication over 5G, by using a combination of the benchmarking tools like $\mathtt{Simu5G}$, $\mathtt{Veins}$~\cite{Sommer2011}, and $\mathtt{PLEXE}$ in $\mathtt{OMNeT++}$. For each of the three \acp{IFT}, we have used three different resource allocation algorithms, \ac{MaxC/I}, \ac{PF}, \ac{DRR}.
    \item Through an exhaustive simulation experiment over varying traffic load (platoon length), we have identified the most suitable \ac{IFT}-resource allocation algorithm combination that meets the corresponding latency and reliability requirements.
    \item We have also carried out exhaustive experiments to identify the most suitable \ac{IFT} that can support multiple platoons in the same bandwidth.
    % \item Through an exhaustive simulation experiment we have identified the most suitable \ac{IFT}-resource allocation algorithm combination that meets the corresponding latency and reliability requirements. 
    % \item We have also carried out exhaustive experiments to identify the most suitable \ac{IFT} that can support multiple platoons in the same bandwidth.
    % \item We have designed a platoon communication mode, we use the benchmarking tool Simu5G~\cite{Nardini2020} to conduct simulations and provide in-depth performance analysis of three resource allocation algorithms: MaxC/I, PF, and DRR.
    % \item We also discuss the limitations of these algorithms under different channel conditions and highlight the need for link adaptation within the resource allocation schemes implemented in the benchmarking tool mentioned above.
\end{itemize}
Our framework is a plug-and-play design where one can change the \ac{IFT}, resource allocation algorithm, number of platoons, and platoon length to carry out the desired experiment in a 5G scenario. Extensive simulation results demonstrate that the combination of the One-Hop \ac{IFT} and \ac{MaxC/I} yields the best \ac{QoS} performance in terms of latency and reliability of the platoons, both in single and multi-platoon scenario. To the best of our knowledge, our paper is the first work that studies the impact of various \acp{IFT} and resource allocation algorithms on the \ac{QoS} performance of vehicle platoons in 5G networks.
% To the best of our knowledge, our paper is the first work that studies the effect of resource allocation on the length and number of vehicle platoons in 5G.

Rest of the paper is organized as follows. Section \ref{sec:SysOver} provides the system overview and the details of the radio resource allocation algorithms. Section \ref{sec:Implementation} outlines the methodology. Section \ref{sec:ComparativeStudy} presents a comparative performance evaluation of three \acp{IFT} and the resource allocation algorithms, followed by a discussion of the advantages and limitations of our model in Section~\ref{sec:Discussion}. Finally, Section \ref{sec:Conclusion} concludes the paper.

\section{System Overview}\label{sec:SysOver}
% In this section, we present our system model.
In this section, we present an overview of our system model, which consists of different (i) platoon \acp{IFT}, and (ii) resource allocation schemes.

\subsection{Platoon \acf{IFT}}\label{sec:PlatoonCommArchs}
Figure~\ref{Fig:PltnArchitectures} shows our reference framework. We have considered a road where up to $M$ platoons (where, $M \in \mathbb{N} =\{1, 2, 3..\}$) are under the coverage of a single 5G \ac{gNB}. The $j^{th}$ platoon contains $N$ vehicles, which is represented as the set $\mathcal{V}_j=\{V_{j,0},V_{j,1},...V_{j,N-1}\}$, where $j \in \{0,1,...,M-1\}$. All vehicles are assumed to have the same length $L$, and the inter-vehicle distance between any two vehicles $V_{j,i}$ and $V_{j,i-1}$ within platoon $\mathcal{V}_j$ is $d$ ($i \in \{0,1,...N-1\}$). In the platoon $\mathcal{V}_j$, the vehicle $V_{j,0}$ is referred to as the \ac{PL}, the vehicles $V_{j,1}$ to $V_{j,N-2}$ are known as \acp{PM}, and the vehicle $V_{j,N-1}$ is called the \ac{PT}. The control model of the platoon vehicles follows \ac{PLF}-based \ac{CACC}~\cite{Segata2014}. Critical vehicle state information (position, speed, and acceleration) within platoon $\mathcal{V}_j$ is exchanged periodically via \acp{CAM} at every $T_p$ seconds. The \ac{CACC} at vehicle $V_{j,i} \in \mathcal{V}_j$ uses its \ac{PL}'s ($V_{j,0}$) and predecessor's ($V_{j,i-1}$) information to periodically update its control input at every $T$ seconds. Here, we have assumed that the packet transmission period is less than the controller update period, i.e., $T_p<T$, such that at every $T$ seconds, vehicle $V_{j,i}$ has both $V_{j,0}$'s and $V_{j,i-1}$'s information for control update.
% and is periodically updated by all vehicles at every $T$ seconds, where $T$ is set to $100 ms$. 
The \acp{CAM} also includes the \emph{lane id} and \emph{platoon id}. The \emph{lane id} indicates the specific lane in which each platoon is traveling, while the \emph{platoon id}, which take values $\{0,1,...M-1\}$, is a unique identifier used to distinguish between the $M$ different platoons. The \emph{platoon id} is exclusive to each of the $M$ platoons, ensuring that all \acp{PM} in a platoon receive the \acp{CAM} disseminated by a \ac{PM} within the same platoon, while vehicles in other platoons discard them. We have considered three \acp{IFT} for \ac{CAM} exchanges in the platoon.
\subsubsection{\textbf{Car-to-Server}} In the Car-to-Server \ac{IFT}, any vehicle $V_{j,i} \in \mathcal{V}_j$ transmits a \ac{CAM} to the \ac{gNB} via an \ac{UL} transmission. The \ac{gNB} then forwards the \ac{CAM} to the recipients in the set ${\mathcal{V}_j\setminus \{V_{j,i}\}}$ via \ac{DL} transmission. Figure~\ref{fig:Car2Server} depicts the Car-to-Server \ac{IFT} for \ac{CAM} transmitted by $V_{j,0}$ to the vehicles in ${\mathcal{V}_j\setminus \{V_{j,0}\}}$ via the \ac{gNB}. \gm{The vehicle $V_{j,i}$'s (Figure~\ref{fig:Car2Server}) controller is updated using its PL's ($V_{j,0}$) and predecessor's ($V_{j,i-1}$) data, which are received via \ac{UL}/\ac{DL} transmission through the base station.}
\subsubsection{\textbf{Multi-Hop}} The Multi-Hop \ac{IFT}, uses \ac{V2V} communication between successive vehicles within platoon $\mathcal{V}_j$. Each vehicle $V_{j,i} \in \mathcal{V}_j$ relays the \ac{CAM} to ensure that all \acp{PM} in the set ${\mathcal{V}_j\setminus \{V_{j,i}\}}$ receive the message. The \ac{CAM} generated by vehicle $V_{j,i}$ is directly transmitted to its immediate successor $V_{j,i+1}$, which then relays the message to the next vehicle $V_{j,i+2}$, and so on, until it reaches the last vehicle $V_{j,N-1}$. Figure~\ref{fig:MultiHop} depicts the Multi-Hop \ac{IFT} for $V_{j,0}$'s \ac{CAM} being relayed to other vehicles in ${\mathcal{V}_j\setminus \{V_{j,0}\}}$. \gm{The vehicle $V_{j,i}$ (Figure~\ref{fig:MultiHop}) receives its PL's ($V_{j,0}$) data via relaying using multiple hops, and predecessor's ($V_{j,i-1}$) data via a single hop to update its controller.}
\subsubsection{\textbf{One-Hop}} In the One-Hop \ac{IFT}, any vehicle $V_{j,i} \in \mathcal{V}_j$ transmits a \ac{CAM} simultaneously to all vehicles in the set ${\mathcal{V}_j\setminus \{V_{j,i}\}}$ within its transmission range. This is done by broadcasting over a shared \ac{IP} address. Only vehicles that have subscribed to this specified \ac{IP} address specific to platoon $\mathcal{V}_j$ will receive the \ac{CAM} from vehicle $V_{j,i}$, others will ignore it. Thus, when $V_{j,i}$ sends a \ac{CAM}, it reaches all the vehicles in the set ${\mathcal{V}_j\setminus \{V_{j,i}\}}$, in a single transmission. Figure~\ref{fig:OneHop} depicts the One-Hop \ac{IFT} for $V_{j,0}$'s \ac{CAM} being simultaneously transmitted to other vehicles in ${\mathcal{V}_j\setminus \{V_{j,0}\}}$. \gm{The vehicle $V_{j,i}$ (Figure~\ref{fig:OneHop}) receives both its PL's ($V_{j,0}$) and predecessor's ($V_{j,i-1}$) data via a single hop to update its controller.} We next discuss different resource allocation algorithms used. 

% The \ac{PL} $V_0$ transmits the \ac{CAM} message simultaneously to all platoon vehicles $V_i$ (where $i = 1, 2, ..., N-1$) within its transmission range. This is done by broadcasting over a shared \ac{IP} address. Only vehicles that have subscribed to this specified \ac{IP} address will receive the \ac{CAM}, others will ignore it. Thus, when $V_0$ sends a \ac{CAM}, it reaches all the vehicles $V_i$ in a single transmission. 
\begin{figure*}[hbt!]
  \includegraphics[width=\textwidth,height=2.5cm]{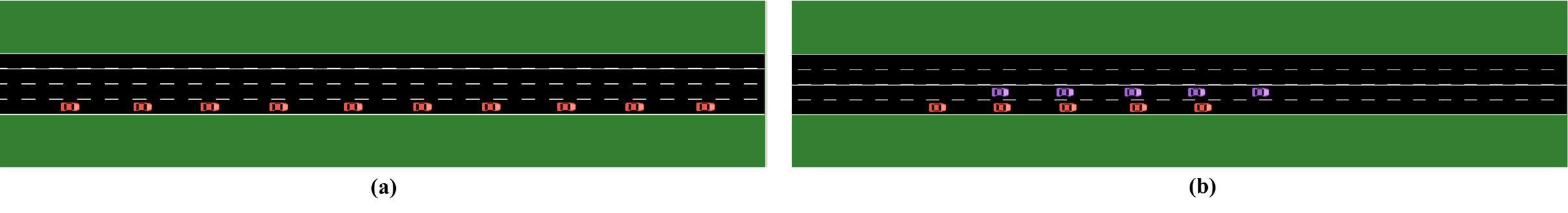}
   \caption{Traffic Simulation in SUMO of (a) single-platoon, and (b) multi-platoons.}
   \label{Fig:SUMO}
\end{figure*}

\subsection{Radio Resource Allocation Schemes}
\label{sec:RAAlgorithms}
The \ac{gNB} is responsible for allocating \acp{RB} for the CAM exchange in each \ac{TTI} of $T_p$ seconds (packet transmission period mentioned in Section~\ref{sec:PlatoonCommArchs}). The resource allocation is executed at the \ac{MAC} layer of \ac{gNB}, and it is subject to the availability of the \acp{RB}. In the frequency domain, a group of $12$ consecutive sub-carriers forms a \ac{RB}. The \ac{SCS} (and hence the \ac{RB} size) is dependent on the numerology $\mu$ and is equal to $2^\mu \times 15$ KHz~\cite{3GPP_R17_2022}. The resource allocation algorithms considered in this work are as follows.
\subsubsection{\textbf{\acf{MaxC/I})}} The \ac{MaxC/I} or the maximum rate resource allocation algorithm aims to maximize the system’s capacity or throughput. This algorithm ranks all the users according to the instantaneous \ac{CQI} values reported by them. It assigns resources to the \acp{UE} having the highest \ac{CQI} value, i.e., \acp{UE} experiencing the best channel conditions \cite{Zain2015}. Thus, the \acp{UE} present in favorable positions, i.e., having the best channel quality, will achieve the high throughput performance. However, system services may be unavailable to the \acp{UE} having worse channel conditions. Hence, \ac{MaxC/I} serves \acp{UE} with the best channel quality conditions at the expense of fairness among them~\cite{Jakimoski2009}.  This resource allocation objective is as follows,
\begin{equation}
    k^* = \underset{k}{\arg\max} R_{k,n},
\end{equation}
where $R_{k,n}$ is the instantaneous transmission rate for UE $k$ over the $n^{th}$ RB. It is calculated by using the Shannon expression for channel capacity at the $t^{th}$ time slot as,
\begin{equation}
    R_{k,n} = \log[1+SINR_{k,n}(t)]
\end{equation}
where, $SINR_{k,n}(t)$ is the Signal-to-Interference Noise Ratio for the $k^{th}$ UE over the $n^{th}$ RB at $t^{th}$ time slot.
\subsubsection{\textbf{\acf{PF}}} The \ac{PF} algorithm provides a balance between fairness and the overall system throughput, i.e., it aims to provide fairness among \acp{UE} while maximizing the throughput~\cite{Zain2015}. The algorithm functions as follows \cite{Barayan2013}; first, the \ac{gNB} obtains the feedback of the instantaneous \ac{CQI} for each \ac{UE} $k$ in time slot $t$ (of $T_p$ seconds duration) in terms of a requested data rate $R_{k,n}(t)$. Then it keeps track of the moving average throughput $T_{k,n}(t)$ for each \ac{UE} $k$ on every RB $n$ within a past window of length $t_c$. The allocation scheme gives priority score $k^*$ to $K$ number of \acp{UE} in $t^{th}$ time slot and RB $n$ that satisfies the maximum relative channel quality condition. The priority score $k^*$ is computed as,
\begin{equation}
    k^* = \underset{k=1,2,..,K}{\arg\max} \frac{[R_{k,n}(t)]^\alpha}{[T_{k,n}(t)]^\beta}
\end{equation}
where $\alpha = 1$ and $\beta = 1$. The \ac{gNB} updates the moving average throughput $T_{k,n}(t)$ of $k^{th}$ \ac{UE} in the consequent time slot as,
\begin{equation}
    T_{k,n}(t+1) = (1 - \frac{1}{t_c})T_{k,n}(t)+\frac{1}{t_c}R_{k,n}(t)
\end{equation}

\subsubsection{\textbf{\acf{DRR}}} This algorithm offers the best fairness among \acp{UE}. It allocates \acp{RB} to \acp{UE} in a cyclic order, without considering the channel conditions \cite{Kuboye2018}. It divides the \ac{V2V} flows into \ac{FIFO} sub-queues and dequeues them from their respective queues in an iterative manner, at each time slot $t$ of $T_p$ seconds duration, in order to execute the \ac{V2V} flows. At each slot $t$, the \ac{DRR} calculates the number of available bits to be transmitted as the sum of the number of allowed bits for transmission and the number of deficit bits from the previous time slot. This provides a higher degree of fairness since it reduces the impact of different packet sizes from various sources. 
\begin{align}
    Q^A_k(t) &= D_k(t-1) + Q^N_k(t) \\
    &b_k(t) \le Q^A_k(t)
\end{align}
Here, for the $k^{th}$ \ac{UE}, $Q^A_k(t)$ is the number of available bits to be transmitted in the $t^{th}$ time slot, $D_k(t-1)$ is the number of deficit bits from $(t-1)^{th}$ time slot, and $Q^N_k(t)$ is the number of number of bits allowed for transmission. The number of bits ($b_k(t)$) that $k^{th}$ \ac{UE} wants to transmit at $t^{th}$ slot is bounded by the available bits $Q^A_k(t)$ computed for the same slot.
\section{Evaluation Methodology}\label{sec:Implementation}
% \subsection{Simulation Setup}
% We simulate vehicle platoon architectures using Simu5G \cite{Nardini2020}, a 5G-enabled library in the OMNeT++ simulation framework. OMNeT++ integrates protocol libraries from the INET Framework \cite{INET2019}. We generate realistic vehicular traffic models using SUMO. SUMO communicates with OMNeT++ via Traffic Control Interface (TraCI), facilitated by Veins framework~\cite{Sommer2011}.  VeinsINET, a sub-project of Veins combines module libraries from Veins \cite{Sommer2011}, INET \cite{INET2019}, and Simu5G \cite{Nardini2020}, enabling end-to-end simulation of C-V2X-based platoon architectures.

% \subsection{Scenario Configuration}
% \label{subsec:ScenConf}
% Our reference scenario (Figure~\ref{Fig:SUMO}) involves a three-lane highway accommodating up to $30$ cars, each of length $L = 5 m$  with an inter-vehicle distance or gap of $d = 11 m$ (including the vehicle length). The study examines $N = 3$ to $10$ vehicles per platoon and $M = 1$ to $3$ platoons. Platoon joining, leaving, or maneuvering is not considered. All platoons 
%  communicate via a single base station with a $200 MHz$ V2X-reserved bandwidth. In our simulation, we vary the number of platoons ($M$) and the platoon length ($N$). The CAM exchange period is set to $T= 30 ms$ with each message size of $110$ bytes (excluding additional headers from lower layers). We perform twenty repetitions, each for $60 s$. Table~\ref{table:Param} summarizes the key simulation parameters for three architectures (Car-to-Server, Multi-Hop, and One-Hop) across both single and multi-platoon setups.
\subsection{Scenario Configuration}
\label{subsec:ScenConf}
To conduct a comprehensive analysis of the different platoon \acp{IFT} under varying platoon lengths $(N)$ and the number of platoons $(M)$ in a 5G network scenario, we have simulated vehicle platoons using $\mathtt{OMNeT++}$, $\mathtt{Veins}$~\cite{Sommer2011}, and $\mathtt{Simu5G}$~\cite{Nardini2020}. First, we have created a highway scenario with three lanes, as shown in Figure~\ref{Fig:SUMO}, using $\mathtt{SUMO}$, $\mathtt{Veins}$ and $\mathtt{PLEXE}$~\cite{Segata2014} to deploy platoon vehicles. The control law of these vehicles follow a \ac{PLF}-based \ac{CACC} strategy. The length of each platoon vehicle, the desired inter-vehicle gap, and the control input update interval has been set to $L=5\ m$, $d=11\ m$ and  
$T = 100\ ms$ respectively. This study does  not consider dynamic change in platoons i.e., platoon joining, leaving, or maneuvering. We have examined two scenarios - (i) single platoon scenario ($M=1$), where the platoon length $(N)$ takes values from $3$ to $10$, and (ii) multiple platoon scenario, where the number of platoons $(M)$ varies as $1, 2$ and $3$. For each of these platoon scenarios, we have considered three \acp{IFT} for the \ac{PL}'s and predecessor's information flow to the \acp{PM} in 5G \ac{eV2X}-based platoon. The three \acp{IFT} are, (i) Car-to-Server, (ii) One-Hop, and (iii) Multi-Hop (refer Section~\ref{sec:SysOver} for details). We have used the $\mathtt{Simu5G}$ framework to enable 5G \ac{eV2X} communication capabilities for the platoon vehicles. In Car-to-Server, all the platoon vehicles (in both the single and multi-platoon scenarios) communicate via a single \ac{gNB}. In One-Hop and Multi-Hop, the platoon vehicles communicate directly with each other via \ac{V2V} links. In all three \acp{IFT}, the \ac{gNB} allocates a transmission bandwidth of 200 MHz for the \ac{CAM} exchange. The \ac{CAM} exchange period and the message size (excluding additional headers from the lower layers) has been set to \ar{$T_p =30\ ms$} and $110$ bytes respectively. We have performed twenty repetitions of each simulation with a simulation time of $60 s$. Table~\ref{table:Param} summarizes the key simulation parameters that we had used for simulating the three \acp{IFT}, across both single and multi-platoon setups.
\begin{table}[htbt!]
\begin{center}
\begin{tabular}{ |p{5cm}|p{2.2cm}| }
  \hline
  \multicolumn{2}{|c|}{\textbf{Communication Parameters}} \\
  \hline
  \centering\textbf{Parameter} & \textbf{Value} \\
  \hline
  Carrier Frequency & 30 GHz \\
  % \hline
  Number of Resource Blocks & 132 \\
  % \hline
  Numerology Index ($\mu$) & 3 \\
  % \hline
  Transmission Bandwidth & 200 MHz \\ 
  % \hline
  Resource Allocation Algorithm & \{\ac{MaxC/I},\ac{PF},\ac{DRR}\} \\ 
  % \hline
  Fixed \ac{CQI} Value & \{3,5,7,9,11\} \\ 
  % \hline
  Packet Transmission Period or \ac{TTI} ($T_p$) & 0.03 s \\ 
  % \hline
  Application Header + Packet Size & 10 + 100 Bytes \\ 
  \hline
  \hline
  \multicolumn{2}{|c|}{\textbf{Simulation Parameters}} \\
  \hline
  \centering\textbf{Parameter} & \textbf{Value} \\
  \hline
  Vehicle Speed & 36 kmph  \\
  % \hline
  Maximum Acceleration & 2.5 $m/s^2$ \\
  % \hline
  Average inter-vehicle spacing ($d$) & 11 m \\
  % \hline
  Vehicle size ($L$) & 5 m \\ 
  % \hline
  Controller Update Period ($T$) & 0.1 s \\
  % \hline
  Maximum number of lanes & 3 \\ 
  % \hline
  Maximum number of platoons ($M$) & 3 \\ 
  % \hline
  Maximum number of cars per platoon ($N$) & 10 \\ 
  % \hline
  Maximum VUE transmit power & 23 dBm \\ 
  % \hline
  Antenna gain & 5 dBi \\
  % \hline
  Vehicle antenna height & 1.6 m \\
  % \hline
  Vehicle receiver noise figure & 13 dB \\
  \hline
\end{tabular}\\ [10pt]
\end{center}
\caption{Parameter Table}
\label{table:Param}
\end{table}

\subsection{Communication Model}
\label{subsec:CommModel}
In all three \acp{IFT} mentioned above, the number of \acp{RB} allocated for \ac{CAM} exchange depends on the transmission bandwidth and the numerology. Platoon vehicles periodically reported their \ac{CQI} values to the \ac{gNB}, which  used this information to select the appropriate \ac{MCS} for the \ac{CAM} transmissions. Further details on \ac{CQI} and corresponding \acp{MCS} are available in~\cite{3GPP2019Rel16}. The $\mathtt{Simu5G}$ tool provides the three resource allocation algorithms: \ac{MaxC/I}, \ac{PF}, and \ac{DRR}, discussed in  Section~\ref{sec:RAAlgorithms}. In this work, we have used these algorithms of the $\mathtt{Simu5G}$ tool, alongside the $\mathtt{Veins}$ and $\mathtt{Plexe}$, to simulate and analyze the performance of each resource allocation method for 5G \ac{eV2X}-based platoon communication. The communication parameters that we have used in our simulations are listed in Table~\ref{table:Param}.

% allocates RBs for the CAM exchanges in each TTI. The resource allocation is executed using a resource allocation algorithm present at the MAC layer of \ac{gNB} and it is subject to the availability of the RBs. The CQI values are reported by vehicles and based on the value, the Modulation and Coding Scheme (MCS) to be used for the CAM transmission is selected by the \ac{gNB}. The different CQI and their corresponding MCSs are mentioned in~\cite{3GPP2019Rel16}. Different CQI effects the transmission range of the CAMs and thus the performance of the resource allocation algorithm for the best-suited architecture for platooning in C-V2X 5G mmWave.

\subsection{Channel Model}
We have simulated the Rayleigh fading channel using a combination of the realistic channel model and the Jakes model~\cite{Liu2011}, which accounts for the Doppler shift in high-mobility vehicular scenarios. $\mathtt{Simu5G}$ provides a realistic channel model based on \ac{3GPP} specifications~\cite{3GPPChannelModel_2017}. In this work, for the platoon scenarios mentioned in Section~\ref{subsec:ScenConf}, we had used the \ac{3GPP} standardized path loss model for 5G \ac{cV2X}~\cite{3GPPPathloss_R15_2019} communication. The path loss $(PL)$ model used for the simulation of \ac{V2V} links is given by,
\begin{equation}
    PL = 32.4 + 20\log_{10}(d) + 20\log_{10}(f_{c}).
\end{equation}
Here, $d$ is the distance between the transmitter and receiver in meters, and $f_{c}$ is the carrier frequency in GHz. The shadow fading  standard deviation is $\sigma\textsubscript{SF} = 3dB$.

% We have integrated a realistic channel model and Jakes model~\cite{Liu2011} to simulate the Rayleigh fading channel in our simulation. Simu5G provides a realistic channel model based on 3GPP specifications~\cite{3GPPChannelModel_2017}. The Jakes model is useful in modeling the Rayleigh fading channel for high-mobility vehicular scenarios as it takes into account the Doppler shift experienced in such a highly dynamic communication channel. 

\begin{figure*}[htbt!]
    \centering
    \begin{subfigure}[b]{0.32\textwidth}
    \centering
    \includegraphics[width=\linewidth,clip]{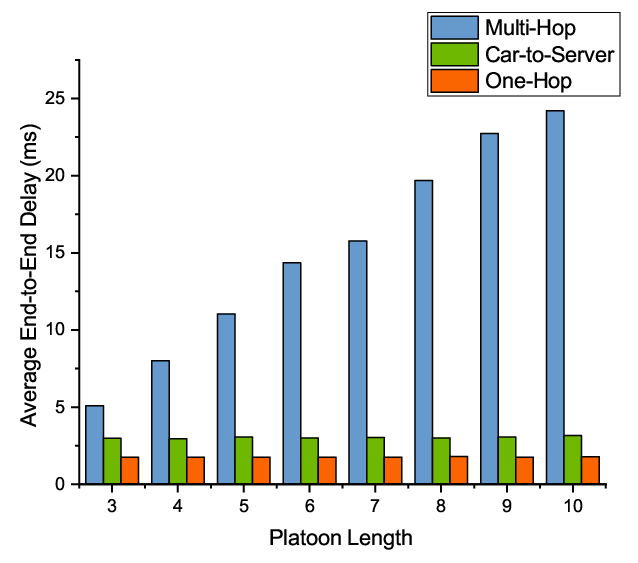}
    \caption{}
    \label{fig:E2EDelay_pltsize}
    \end{subfigure}
    \hfill % Add horizontal space between figures
    \begin{subfigure}[b]{0.32\textwidth}
    \centering
    \includegraphics[width=\linewidth,clip]{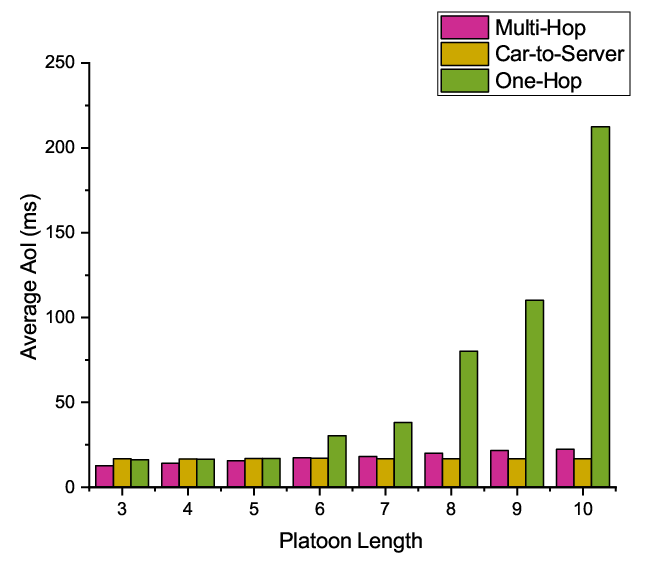}
    \caption{}
    \label{fig:AoI_pltsize}
    \end{subfigure}
    \hfill % Add horizontal space between figures
    \begin{subfigure}[b]{0.32\textwidth}
    \centering
    \includegraphics[width=\linewidth,clip]{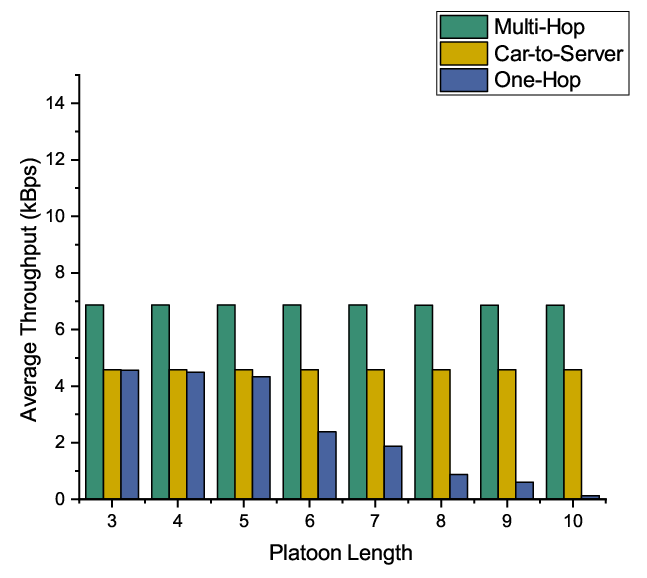}
    \caption{}
    \label{fig:Throughput_pltsize}
    \end{subfigure}
    % \vspace{-0.4cm}
    \caption{Effect of increasing the number of vehicles $(N)$ in a Single platoon $(M=1)$ on the average (a) End-to-End (E2E) delay, (b) Age of Information (AoI), and (c) Throughput observed for PL-to-PT CAM exchanges across the three platoon \acp{IFT}.}
    \label{fig:Inc_Len_Platoon}
    % \vspace{-0.7cm}
\end{figure*}
\begin{figure*}[htbt!]
    \centering
    \begin{subfigure}[b]{0.32\textwidth}
    \centering
    \includegraphics[width=\linewidth,clip]{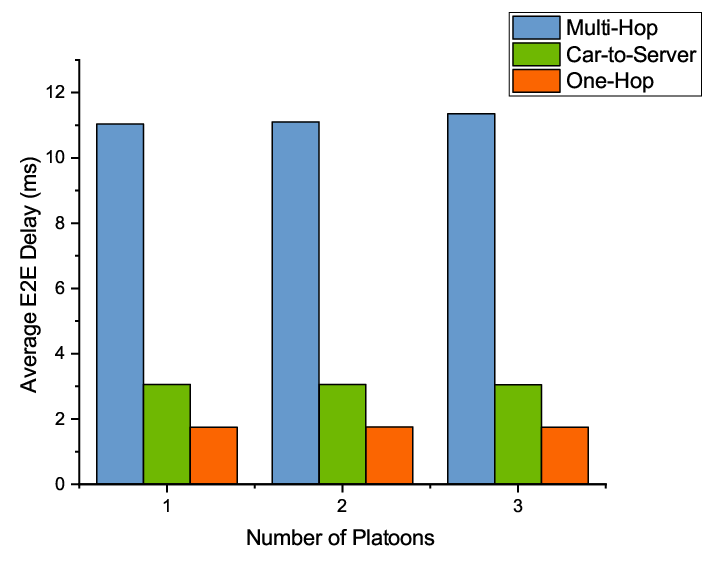}
    \caption{}
    \label{fig:E2E_PltNum}
    \end{subfigure}
    \hfill % Add horizontal space between figures
    \begin{subfigure}[b]{0.32\textwidth}
    \centering
    \includegraphics[width=\linewidth,clip]{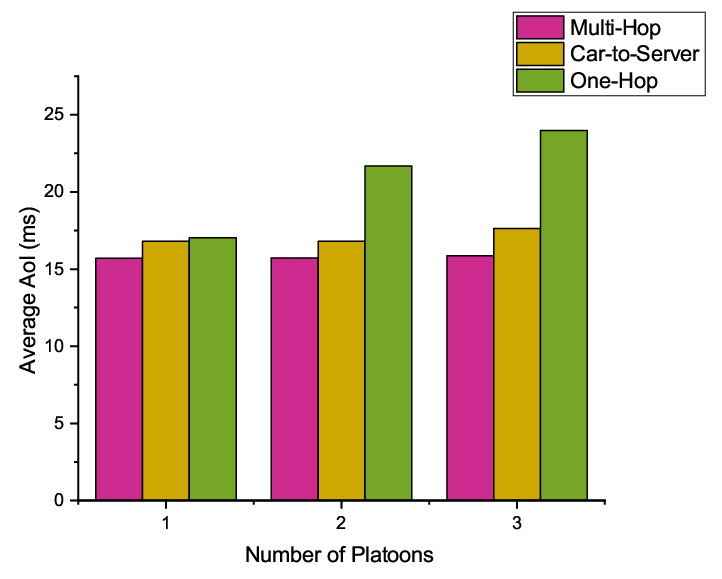}
    \caption{}
    \label{fig:AoI_pltNum}
    \end{subfigure}
    \hfill % Add horizontal space between figures
    \begin{subfigure}[b]{0.32\textwidth}
    \centering
    \includegraphics[width=\linewidth,clip]{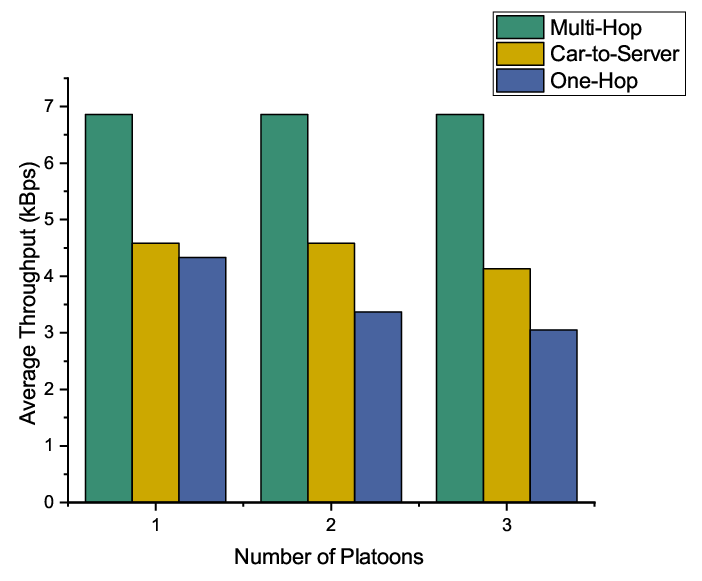}
    \caption{}
    \label{fig:Throughput_pltNum}
    \end{subfigure}
    % \vspace{-0.4cm}
    \caption{Effect of increasing the number of platoons $(M)$ for a fixed platoon size $(N=5)$ on the average (a) \acf{E2E} delay, (b) \acf{AoI} and (c) Throughput observed for PL-to-PT CAM exchanges across the three platoon \acp{IFT}.}
    \label{fig:Inc_Num_Platoons}
    % \vspace{-0.7cm}
\end{figure*}

\section{Performance Evaluation}\label{sec:ComparativeStudy}
We have compared the different platoon \acp{IFT}: Car-to-Server, Multi-Hop, and One-Hop, in terms of \ac{QoS} parameters, such as average \ac{E2E} delay, \ac{AoI}, and throughput, are analyzed for varying platoon lengths and number of platoons. To find the most suitable \ac{IFT}-resource allocation algorithm combination, we have evaluated the performance of the three resource allocation strategies: \ac{MaxC/I}, \ac{PF}, and \ac{DRR} by using the best-performing platoon \ac{IFT} in our experiments. The different \ac{QoS} parameters used are defined next.
\subsection{QoS Parameters}\label{AA}
\begin{itemize}
    \item  \textbf{Average \acf{E2E} Delay}: Measured in milliseconds, this metric refers to the time taken for a packet to travel from source to destination across the network, averaged over the total number of packets transmitted.
    \item \textbf{Average \acf{AoI}}: This metric is used to measure the freshness of information. It represents the time elapsed between the generation of an update packet at the transmitter and the successful reception of the previous update packet at the receiver. It is also measured in milliseconds.
    \item \textbf{Average Throughput}: This is the total amount of data successfully received, calculated as the product of the number of successfully received packets and the packet size,  divided by the total simulation time. It is measured in kilobytes per second (kBps).
    \item \textbf{Reception Probability}: This is measured in terms of the ratio of successfully received messages to the total number of messages transmitted.
\end{itemize}
% \noindent\\
% \\
% \\
% % It is the ratio of the product of the number of successfully received packets and the size of the packet to the total duration of simulation time. It is measured in kilobytes per second.\\
\subsection{Impact of Vehicular Traffic Load}\label{AA}
\subsubsection{\textbf{Single Platoon}}\label{AA}
% Figure~\ref{fig:Inc_Len_Platoon} presents the results for 
In the first set of results, we observe the performance of the three \acp{IFT} (Multi-Hop, Car-to-Server, and One-Hop) under varying traffic loads, focusing on a single platoon ($M=1$) with increasing platoon length $N$ ranging from $3$ to $10$ vehicles. For this experiment, we fix the resource allocation algorithm to \ac{MaxC/I}. The results depicted in Figure~\ref{fig:Inc_Len_Platoon} highlight the performance of the \ac{PL}-to-\ac{PT} CAM exchanges for different \acp{IFT} within a single platoon.
Figures~\ref{fig:E2EDelay_pltsize},~\ref{fig:AoI_pltsize}, and~\ref{fig:Throughput_pltsize} shows the variations in the \ac{QoS} parameters, such as average \ac{E2E} delay, \ac{AoI}, and throughput, as the platoon size increases. The average \ac{E2E} delay increases with the increase in platoon size, suggesting larger platoons face more delays in data exchanges. In the Multi-Hop \ac{IFT}, relaying messages from PL to PT results in the highest E2E latency, making it the least preferred for low-latency communication. In contrast, One-Hop \ac{IFT} exhibits the lowest E2E delay and is the most suitable \ac{IFT} for real-time safety-critical platoon applications. An increasing trend is observed for average \ac{AoI}, indicating that the freshness of data received decreases with larger platoon size. The average throughput decreases with increasing platoon size. These observations highlight the challenges of maintaining low latency and high-reliability communication in platoon scenarios with more number of vehicles.
\subsubsection{\textbf{Multiple Platoon}}\label{AA}
The second set of results is obtained for the performance of the same three \acp{IFT} (Multi-Hop, Car-to-Server, and One-Hop), but this time with varying numbers of platoons $M=\{1,2,3\}$, and keeping the platoon size fixed at $N=5$. Similar to the single platoon scenario, here also we fix the resource allocation algorithm to \ac{MaxC/I}. The results (Figure~\ref{fig:Inc_Num_Platoons}) illustrate the performance of the \ac{PL}-to-\ac{PT} CAM exchanges in a multi-platoon scenario for $M=1, M=2$, and $M=3$.
% As shown in Figures~\ref{fig:E2E_PltNum},~\ref{fig:AoI_pltNum}, and~\ref{fig:Throughput_pltNum}, 
As shown in Figure~\ref{fig:E2E_PltNum}, the average \ac{E2E} delay increases with the increasing number of platoons. Notably, even in the multi-platoon scenarios, the Multi-Hop \ac{IFT} exhibits the highest E2E delay, while the One-Hop \ac{IFT} demonstrates the lowest delay. This trend indicates that managing communication among multiple platoons simultaneously causes delayed data exchanges. Additionally, the average \ac{AoI} (Figure~\ref{fig:AoI_pltNum}) increases with $M$, which signifies the reduced data freshness as more platoons are involved. Furthermore, the average throughput (Figure~\ref{fig:Throughput_pltNum}) declines with the addition of more platoons. These results highlight the complexity of maintaining low latency and high-reliability communication in multiple platoon scenarios as the traffic load increases with the increasing number of platoons $M$.
Overall, our results suggest that One-Hop is the most suitable \ac{IFT} for effective communication for both single platoon and multiple platoon scenarios. A detailed discussion of this deduction is presented in Section~\ref{sec:Discussion}.

\begin{figure*}[tbp]
    \centering
    \begin{subfigure}[b]{0.32\textwidth}
    \centering
    \includegraphics[width=\linewidth,clip]{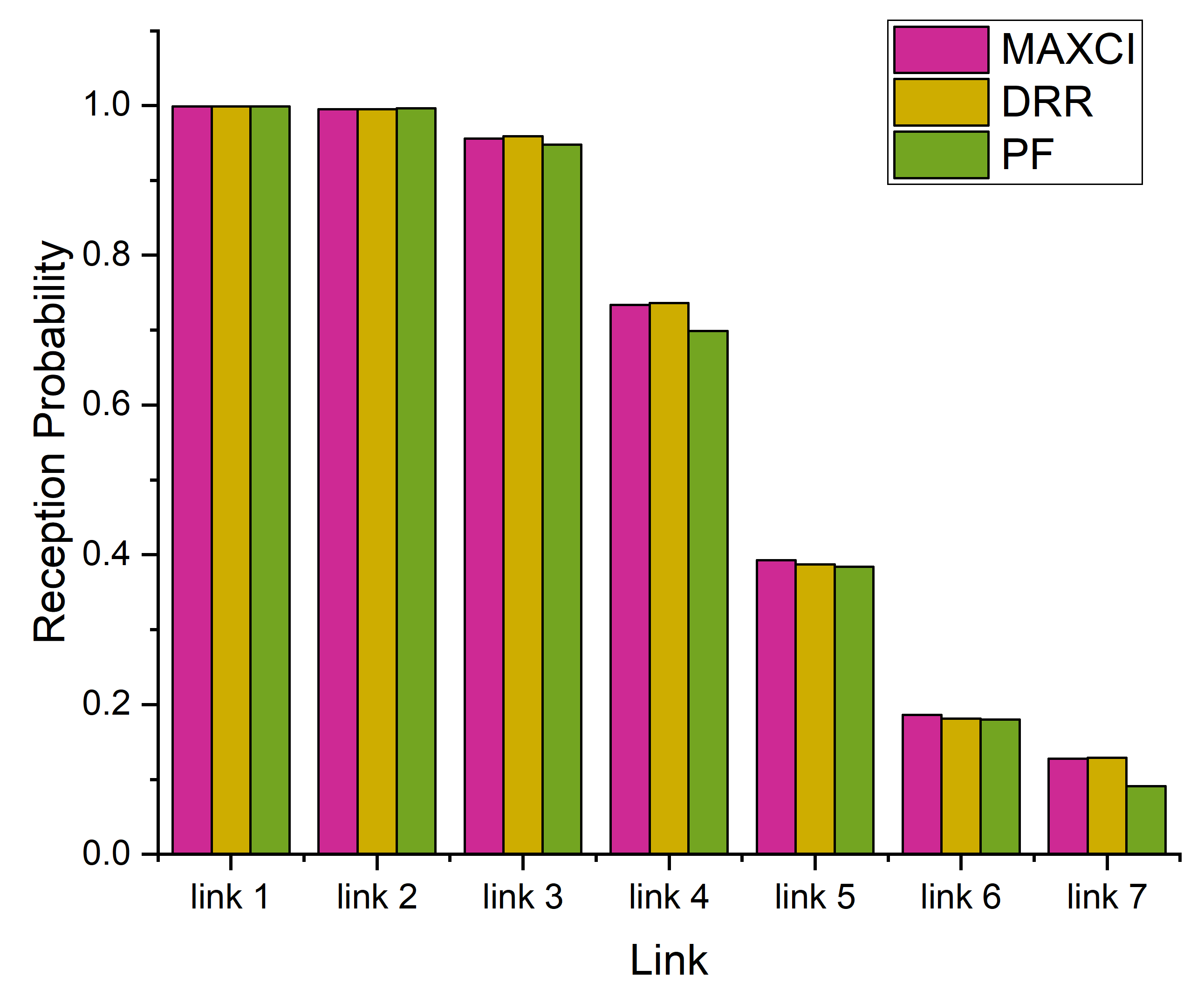}
    \caption{}
    \label{fig:RP_OH_881_CQI7}
    \end{subfigure}
    \hfill % Add horizontal space between figures
    \begin{subfigure}[b]{0.32\textwidth}
    \centering
    \includegraphics[width=\linewidth,clip]{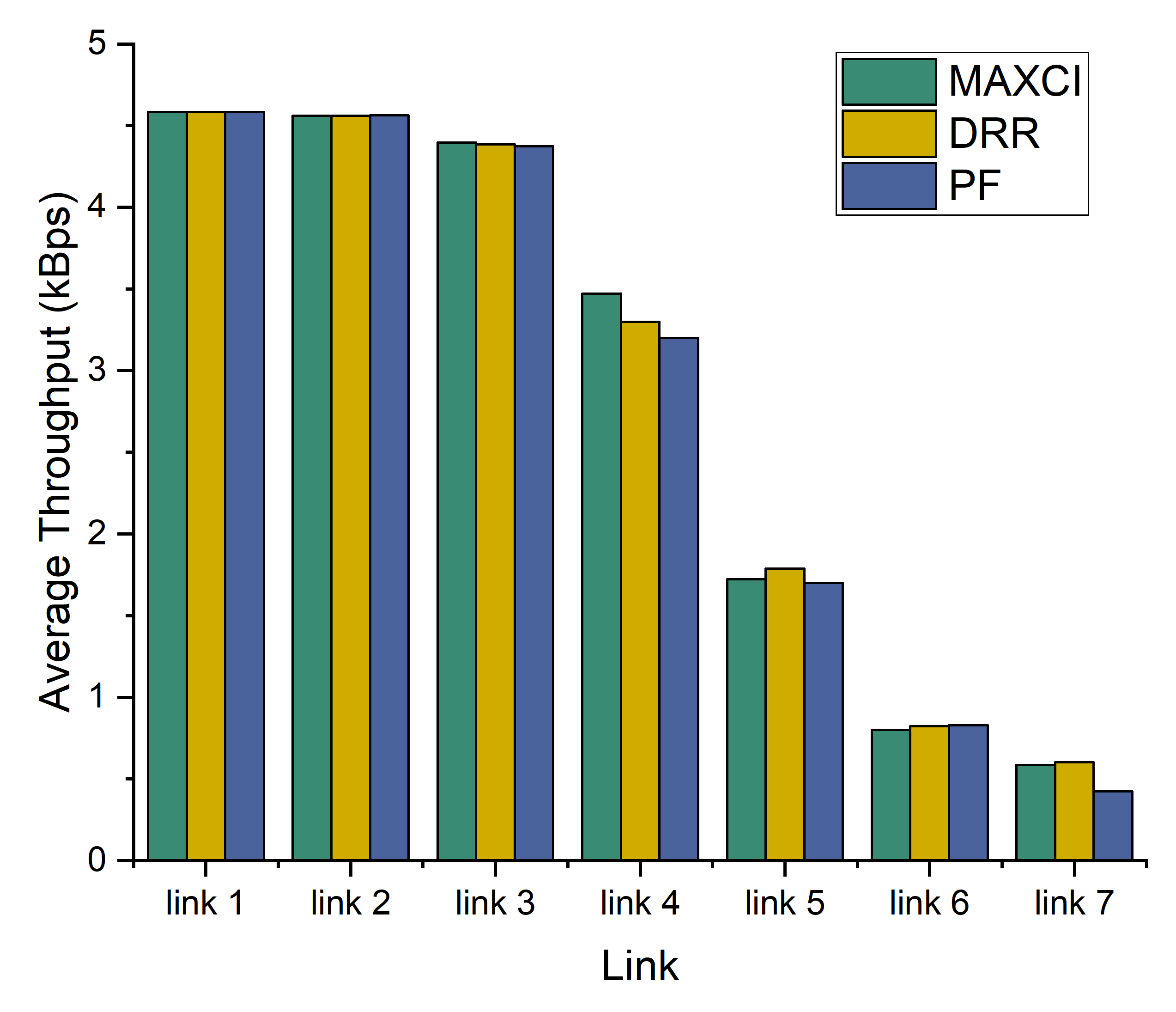}
    \caption{}
    \label{fig:Thrghpt_OH_881}
    \end{subfigure}
    \hfill % Add horizontal space between figures
    \begin{subfigure}[b]{0.32\textwidth}
    \centering
    \includegraphics[width=\linewidth,clip]{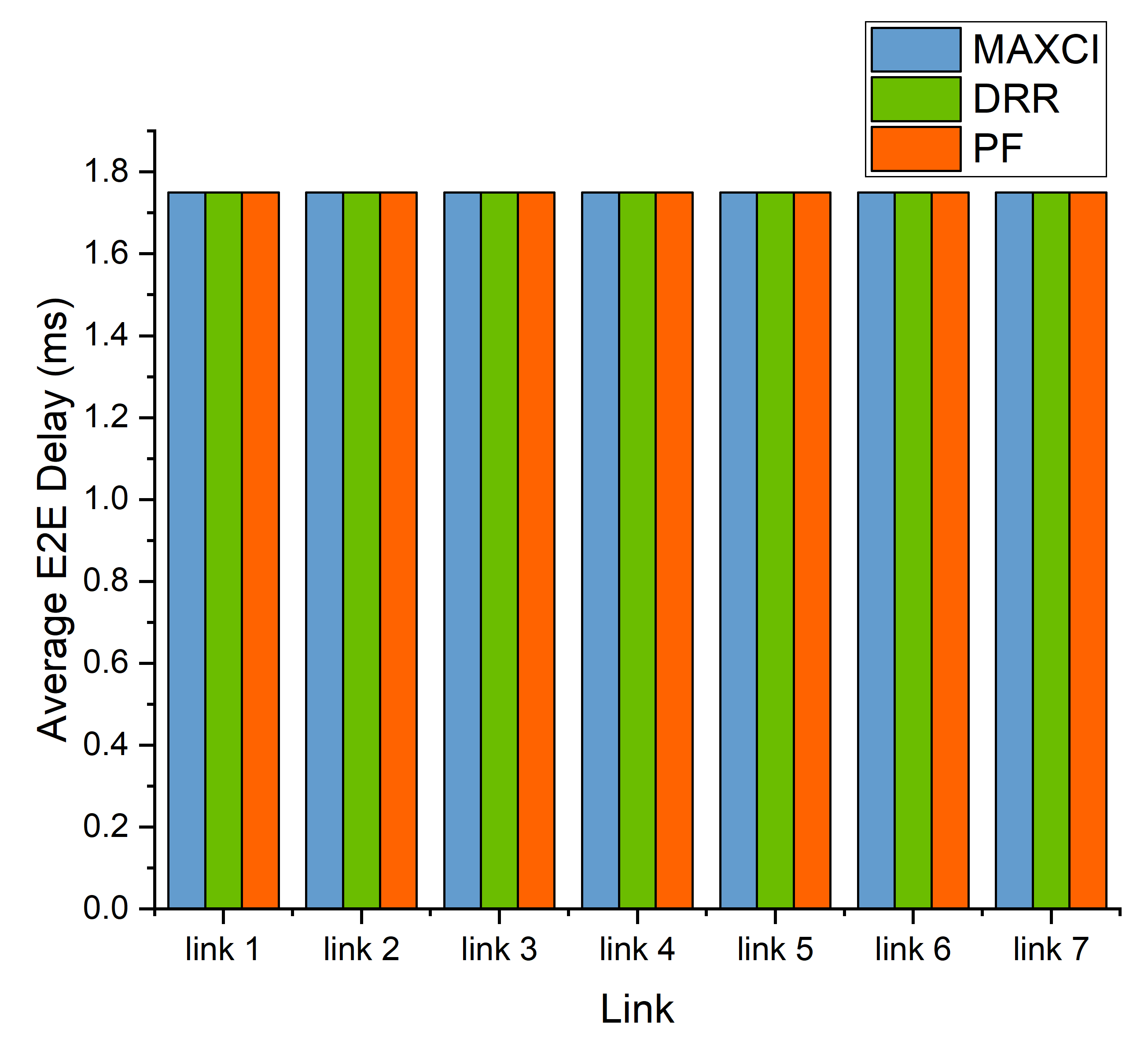}
    \caption{}
    \label{fig:Delay_OH_881_CQI7}
    \end{subfigure}
    % \vspace{-0.4cm}
    \caption{Effect of across three resource allocation algorithms on each D2D link of a single One-Hop \ac{IFT}-based platoon $(M=1)$ of size $(N=8)$ on (a) Reception Probability, (b) Average Throughput, and  (c) Average \acf{E2E} Delay}
    \label{fig:RP_AvgThr_Delay_link}
    % \vspace{-0.7cm}
\end{figure*}

\subsection{Impact of Resource Allocation Schemes}\label{AA}
In this section, at first, we assess the performance of the \ac{MaxC/I} algorithm for the communication between the \ac{PL} and the \ac{PT} for the One-Hop \ac{IFT}-based platoon under varying channel conditions (\acp{CQI}) and lengths of a single platoon. This study helps analyze how the \ac{PL}-to-\ac{PT} link in One-Hop \ac{IFT} experiences different channel conditions due to differences in transmission/reception distances among platoon vehicles. Second, we consider the same One-Hop \ac{IFT}-based platoon of fixed length $N=8$, for a fixed \ac{CQI} setting of $7$, for our study to evaluate the performance of all three resource allocation algorithms (\ac{MaxC/I}, \ac{PF}, and \ac{DRR}) for each D2D link. Our major interest is to identify the best-performing resource allocation algorithm for One-Hop \ac{IFT}, from among three resource allocation algorithms (\ac{MaxC/I}, \ac{PF}, and \ac{DRR}) under different channel conditions, and consequently different \ac{MCS} (and \ac{CQI}) settings of 5G (\ac{MCS} and \ac{CQI} are discussed in Section~\ref{subsec:CommModel}).

% we consider a One-Hop \ac{IFT}-based platoon of length $N=8$ for a fixed \ac{CQI} setting of $7$ for our study to evaluate the performance of all three resource allocation algorithms (\ac{MaxC/I}, \ac{PF}, and \ac{DRR}) for each D2D link (Figure~\ref{fig:RP_AvgThr_Delay_link}). Second, we assess the performance of the \ac{MaxC/I} algorithm for the communication between the \ac{PL} and the \ac{PT} for the One-Hop \ac{IFT} under varying channel conditions (\acp{CQI}) and lengths of a single platoon. This study helps analyze how the \ac{PL}-to-\ac{PT} link in One-Hop \ac{IFT} experiences different channel conditions due to differences in transmission/reception distances among platoon vehicles. Our major interest is to identify the best-performing resource allocation algorithm for One-Hop \ac{IFT}, from among three resource allocation algorithms (\ac{MaxC/I}, \ac{PF}, and \ac{DRR}) under different channel conditions, and consequently different \ac{MCS} (and \ac{CQI}) settings of 5G (\ac{MCS} and \ac{CQI} are discussed in Section~\ref{subsec:CommModel}).

\begin{figure}[htbt!]
    \centering
    \begin{subfigure}[b]{0.24\textwidth}
    \centering
    \includegraphics[width=\linewidth,clip]{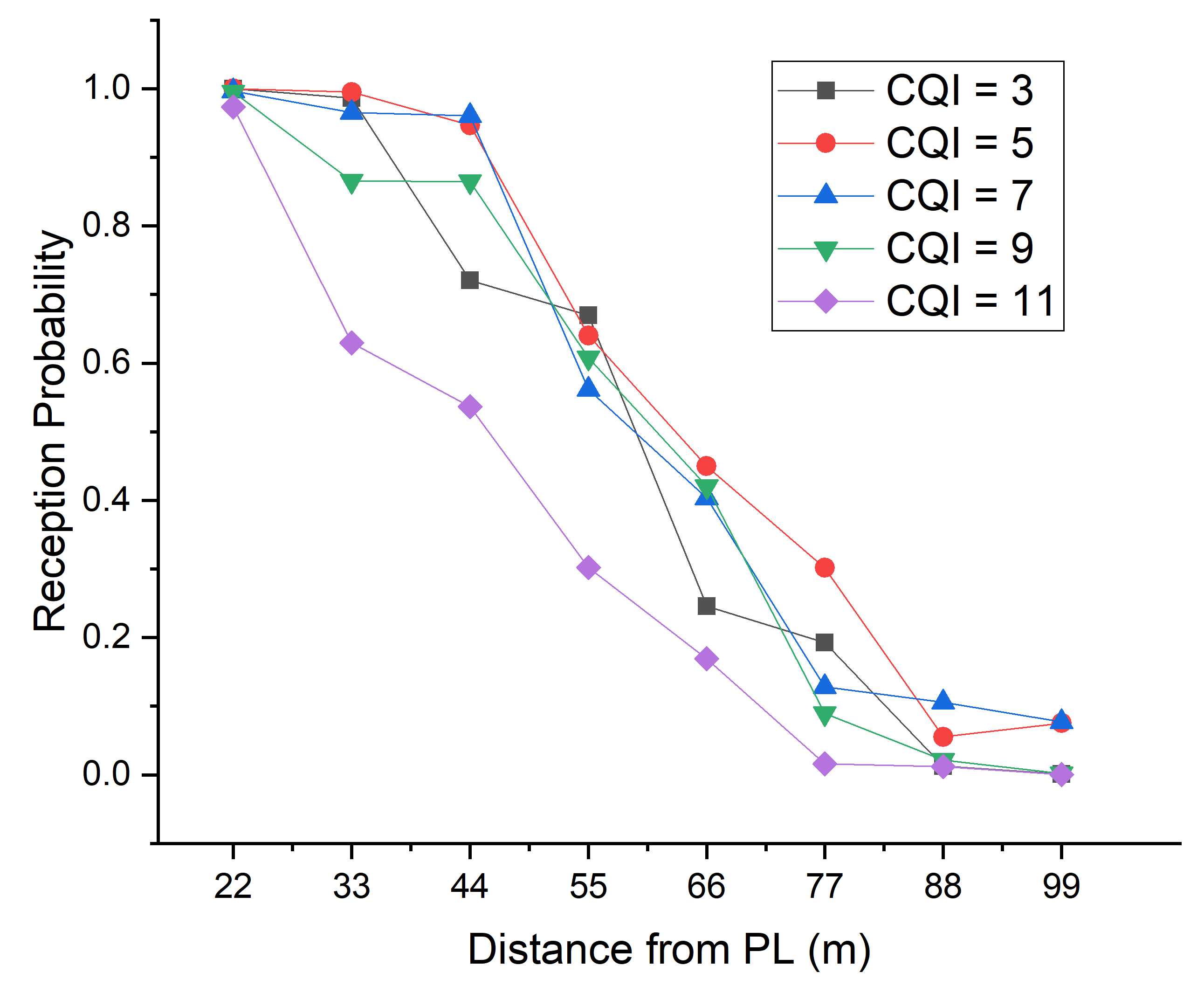}
    \caption{}
    \label{fig:Recep_Prob_MAXCI}
    \end{subfigure}
    \hfill % Add horizontal space between figures
    \begin{subfigure}[b]{0.24\textwidth}
    \centering
    \includegraphics[width=\linewidth,clip]{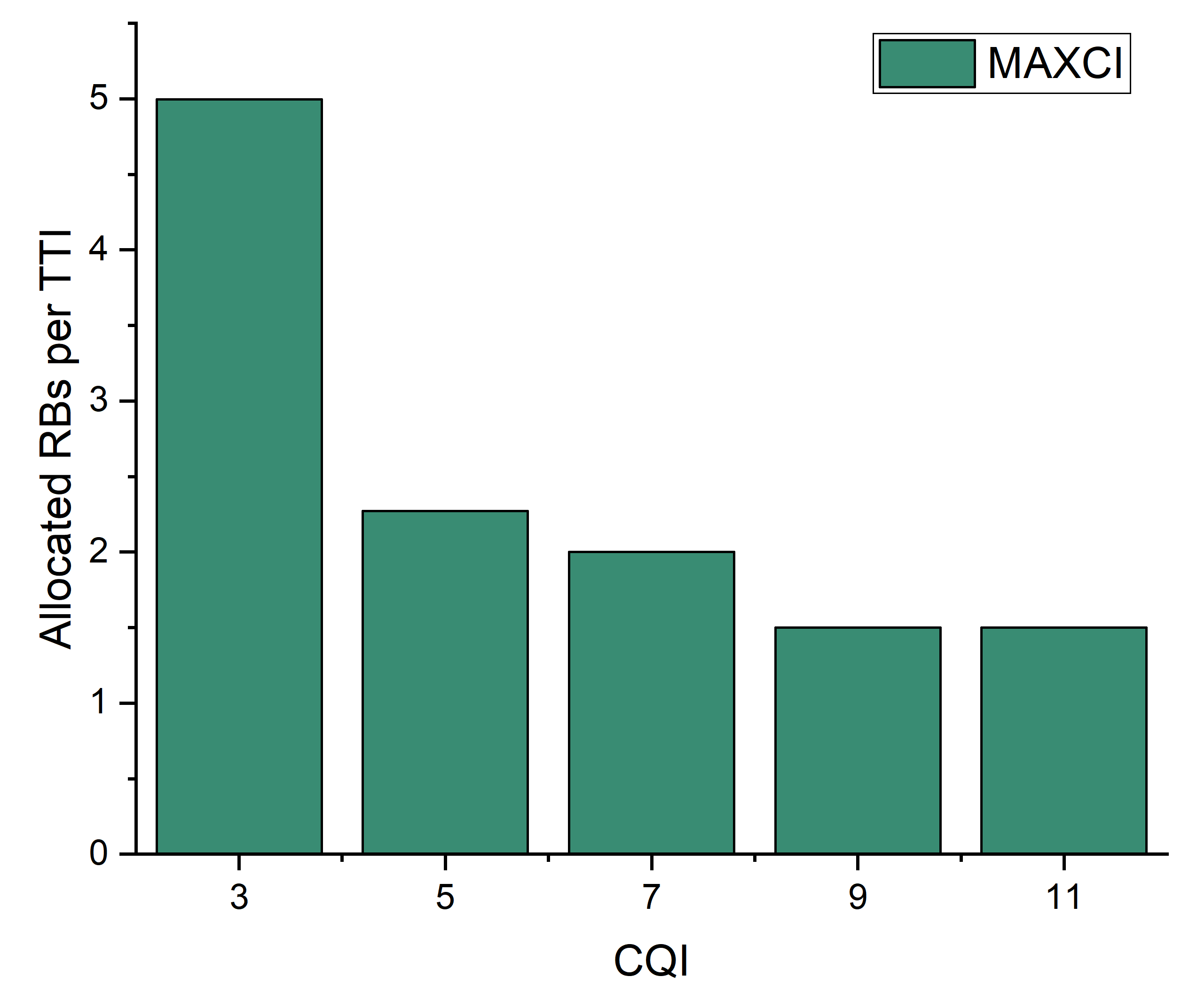}
    \caption{}
    \label{fig:RB_MAXCI}
    \end{subfigure}
    % \vspace{-0.4cm}
    \caption{(a) Effect of increase in distance between \ac{PT} and \ac{PL} (platoon length increases from $N=3$ to $10$) on the reception probability at the PT for \ac{MaxC/I}, (b) Allocated \acp{RB} per \ac{TTI} for different \acp{CQI} using \ac{MaxC/I} for platoon length $N=5$.}
    \label{fig:RP_RB_MXCI}
    % \vspace{-0.5cm}
\end{figure}

Figure~\ref{fig:Recep_Prob_MAXCI} reports the probability of successful reception of the \acp{CAM} from the \ac{PL} (transmitter) to the \ac{PT} (receiver), using the \ac{MaxC/I} resource allocation algorithm for different \ac{CQI} settings. The link between the \ac{PL} and the \ac{PT} is referred to as the worst-case link for a given platoon length $N$, as \ac{PT} is the farthest vehicle from the \ac{PL}, and therefore it experiences the worst signal strength from the \ac{PL} as compared to other \acp{PM}. The $x$-axis of the plot indicates the distance between the \ac{PL} and the \ac{PT} for different platoon lengths $N$, for example, \ac{PT}'s distance from \ac{PL} is $44 m$ for a platoon with $N=5$ vehicles with an intra-platoon gap of $11 m$ (the gap is fixed). As observed from the plot, the value of this metric decreases as the platoon length (\ac{PT}'s distance from the \ac{PL}) increases. This happens due to reduced received signal strength as the platoon length (distance between transmitter and receiver) increases. Compared to lower \ac{CQI} settings, the higher \ac{CQI} settings suffer from a lower probability of reception due to higher error probabilities associated with their corresponding \ac{MCS}. Platoon of length $N=5$ (distance is $44m$) reports a reception probability of at least $0.7$ for lower \ac{CQI} settings. Notably, the \ac{CQI} settings $5$ and $7$ report the highest reception probabilities, close to $1$, ensuring high reliability at the \ac{PT}.

Figure~\ref{fig:RB_MAXCI} shows the number of \acp{RB} allocated per \ac{TTI} using the \ac{MaxC/I} algorithm the platoon length is $N=5$. The plot clearly shows that the employed \ac{MCS} setting affects the number of allocated \acp{RB}. Lower \ac{CQI} settings require more \acp{RB}, resulting in higher transmission overhead. For the successful reception of \acp{CAM}, all the \acp{RB} must be correctly decoded. As seen in Figure~\ref{fig:Recep_Prob_MAXCI}, the \ac{CQI} of setting $3$ has a higher probability of erroneous reception, leading to a lower reception probability, as compared to \ac{CQI} $5$ and $7$.

Figures~\ref{fig:RP_OH_881_CQI7},~\ref{fig:Thrghpt_OH_881}, and~\ref{fig:Delay_OH_881_CQI7} report the reception probability, average throughput, and average \ac{E2E} delay of the \acp{CAM} transmitted from \ac{PL} over each link in a One-Hop \ac{IFT}-based platoon of length $N=8$, and at CQI $7$, for the MaxC/I, \ac{PF} and \ac{DRR} resource allocation algorithms. There is a decreasing trend in the reception probability and average throughput as the distance increases, especially for the worst-case link (the link between the PL and the PT). The \ac{MaxC/I} achieves relatively higher reception probability and average throughput in comparison to \ac{PF} and \ac{DRR} for links $1$ to $4$ (links with favourable channel conditions). This trend gets disrupted for links beyond $4$ (links with unfavourable channel conditions). \ar{The \ac{E2E} delay for all three resource allocation algorithms remains constant over all the links as they communicate via direct One-Hop. The resource allocation schemes (\ac{MaxC/I}, \ac{PF}, and \ac{DRR}) under consideration in this study mainly influence the throughput and the fairness (refer section~\ref{sec:RAAlgorithms}) with which each platoon vehicle is served, and hence the variation in \ac{E2E} delay is negligible.}

% The algorithms perform similarly across the links with small variations as we move towards the worst-case link. The E2E delay for all three resource allocation algorithms remains constant over all the links.
\section{Discussion and Limitations}\label{sec:Discussion}
Table~\ref{table:SimResults}, which shows the worst-case results for all scenarios, presents a comparative analysis of the three communication \ac{IFT}s: Multi-Hop, Car-to-Server, and One-Hop, for a platoon length of $N=5$. It reveals important trends in both single and multi-platoon scenarios. In both cases, the One-Hop \ac{IFT} consistently has the lowest E2E delay, making it the most suitable communication \ac{IFT} for real-time safety-critical platoon applications as it meets the 3GPP Release 16 latency requirements~\cite{3GPP2019Rel16}. The lower average \ac{AoI} of this \ac{IFT}  indicates fresher and more up-to-date data, which is crucial for the safety of platoon vehicles. Although its throughput is slightly lower than other \ac{IFT}s, the minimal delay and \ac{AoI} give One-Hop a clear advantage. In contrast, the Multi-Hop \ac{IFT} has the highest \ac{E2E} delay, making it the least preferred \ac{IFT}, despite its lower \ac{AoI}. Meanwhile, the Car-to-Server \ac{IFT} offers a decent balance between \ac{AoI} and throughput but falls short in latency when compared to One-Hop. Thus, by prioritizing low latency and data freshness, One-Hop \ac{IFT} emerges as the most favorable information flow strategy for both single and multi-platoon scenarios.   
\begin{table}[htbt!]
\begin{center}
 \begin{tabular}{ |p{2cm}|p{1.5cm}|p{1.5cm}|p{1.4cm}| }
 \hline
 \multicolumn{4}{|c|}{\textbf{Multi-platoon for $N= 5$}} \\
 \hline
 \textbf{Platoon \ac{IFT}s} & \textbf{Average E2E Delay (ms)} & \textbf{Average AoI (ms)} & \textbf{Average Through-
 put (kBps)}\\
 \hline 
 Multi-Hop & 11.35 & 15.87 & 6.86\\
 % \hline
 Car-to-Server & 3.6 & 17.62 & 4.58\\
 % \hline
 One-Hop & 1.75 & 23.99 & 4.33\\
 \hline
 \hline
 \multicolumn{4}{|c|}{\textbf{Single-platoons for $N= 5$}} \\
 \hline
 \textbf{Platoon \ac{IFT}s} & \textbf{Average E2E Delay (ms)} & \textbf{Average AoI (ms)} & \textbf{Average Through-
 put (kBps)}\\
 \hline 
 Multi-Hop & 11 & 15.7 & 6.86\\
 % \hline
 Car-to-Server & 3.3 & 16.92 & 4.58\\
 % \hline
 One-Hop & 1.75 & 17.03 & 4.33\\
 \hline
\end{tabular}
% \vspace{0.2cm}
\caption{Comparison of the QoS parameters of single $(M=1)$ and multi-platoons $(M={1,2,3})$ of platoon length $(N=5)$ for three \acp{IFT}.}
\label{table:SimResults}
\end{center}
\end{table}

\begin{wrapfigure}[16]{l}{0.6\columnwidth}
% \captionsetup{justification=raggedright}
    \centering    \includegraphics[width=0.65\columnwidth,trim={0cm 0cm 0cm 2cm}]{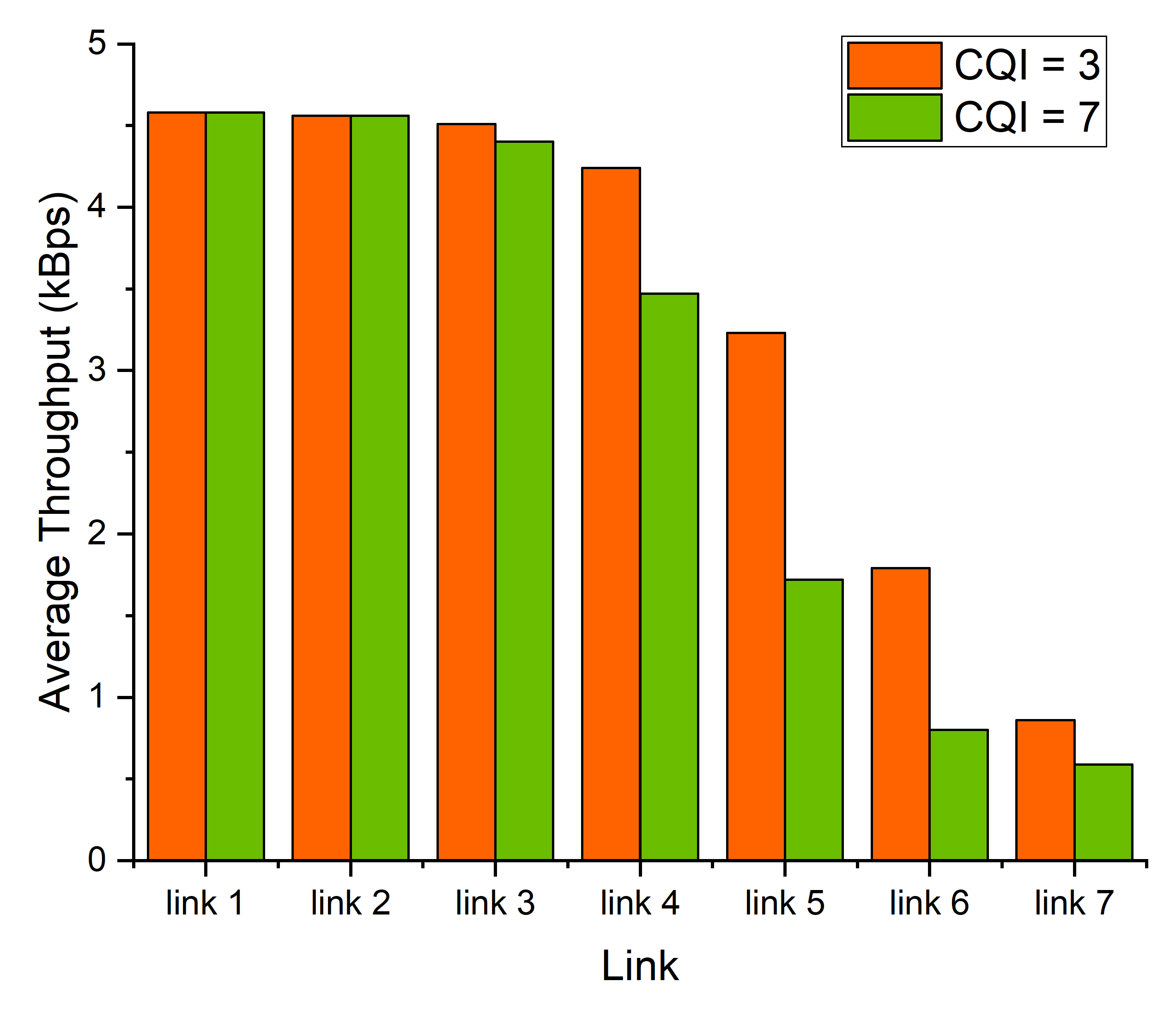}
    \caption{Average Throughput at each D2D link for \ac{MaxC/I} at CQI 3 and 7 for fixed platoon length of $N=8$.}
    \label{fig:AvgThrCQI3_7}
\end{wrapfigure}

The One-Hop \ac{IFT} provided by the authors of $\mathtt{Simu5G}$ in \cite{Nardini2016} is under the constraint that only fixed \ac{CQI} mode can be used, i.e., the provision of channel feedback-based \ac{MCS} selection for link adaptation is not provided. A comparison of the performance of \ac{CQI} mode $3$ and $7$ for \ac{MaxC/I} algorithm and fixed platoon length of $N=8$, as illustrated in Figure~\ref{fig:AvgThrCQI3_7}, explains the limitation of using fixed \ac{CQI}. Since the channel feedback is not present, fixed \ac{CQI} of $7$ and $3$ are being used even when the channel condition is degrading as we move towards the worst-case links in the platoon. In the case of \ac{CQI} $7$, the number of \acp{RB} allocated for transmission is less compared to CQI $3$, as we have seen in Figur~\ref{fig:RB_MAXCI}. So in the worst case links, the requirement of the number of \acp{RB} is more. However, we are allocating less number of \acp{RB} for \ac{CQI} $7$. This leads to packet drops and hence the degradation in the average throughput. \ac{CQI} $3$ performs better as we move towards the worst case link as it allocates more number of \acp{RB}, that meet the throughput requirements of the link.
\section{Conclusion}\label{sec:Conclusion}
In this work, we have developed an end-to-end system model for 5G \ac{eV2X}-based vehicle platooning considering different platoon \ac{IFT}s like Car-to-Server, One-Hop, and Multi-Hop, using the benchmarking tools $\mathtt{Simu5G}$, $\mathtt{Veins}$, and $\mathtt{PLEXE}$ in $\mathtt{OMNeT++}$. We have first evaluated the performance of the three platoon communication \ac{IFT}s, for varying platoon lengths and number of platoons, against the stringent latency and reliability standards of 5G \ac{eV2X}. From the results, we have determined One-Hop to be the most-suitable platoon \ac{IFT} that achieves desirable latency and reliability performance. Then, through further simulations, we have analyzed the performance of the different resource allocation algorithms, namely, \ac{MaxC/I}, \ac{PF}, and \ac{DRR}, and identified that for One-Hop (most-suitable \ac{IFT}), the \ac{MaxC/I} performs the best among the others when the channel conditions are most favourable. However, the limitations of the \ac{MaxC/I} algorithm, particularly for fixed \ac{CQI} value in the $\mathtt{Simu5G}$ benchmarking tool, pose challenges in the worst-case link scenario within One-Hop-based platooning. For future work, we plan to develop a link-adaptation-based resource allocation strategy for the $\mathtt{Simu5G}$ tool. This strategy will dynamically allocate resources to platoon vehicles based on varying channel conditions (\ac{CQI}) to address the performance degradation observed in worst-case link scenarios, further improving the robustness of One-Hop-based platooning communication.

\begin{acronym}
	\acro{2G}{2$^\text{nd}$ Generation}
	\acro{3G}{3$^\text{rd}$ Generation}
	\acro{4G}{4$^\text{th}$ Generation}
	\acro{5G}{5$^\text{th}$ Generation}
	\acro{3GPP}{$\text{3}^\text{rd}$ Generation Partnership Project}
	\acro{A3C}{Actor-Critic}
	\acro{ABR}{Adaptive Bitrate}
	\acro{BS}{Base Station}
	\acro{BUE}{Best-Effort UE}
	\acro{BWP}{Bandwidth Partition}
	\acro{CDN}{Content Distribution Network}
	\acro{CDF}{Cumulative Distribution Function}
	\acro{CSI}{Channel State Information}
	\acro{CUE}{Cellular User Equipment}
        \acro{cV2X}{Cellular Vehicle to Everything}
	\acro{DASH}{Dynamic Adaptive Streaming over HTTP}
 \acro{DSRC}{Direct Short Range Communication}
	\acro{DL}{deep learning}
 \acro{DRL}{Deep Reinforcement Learning}
	\acro{DRX}{Discontinuous Reception}
	\acro{D2D}{Device-to-Device}
	\acro{EDGE}{Enhanced Data Rates for \ac{GSM} Evolution.}
			\acro{gNB}{general NodeB}
	\acro{eNB}{evolved NodeB}
	\acro{GSM}{Global System for Mobile}
	\acro{FL}{Federated Learning}
        \acro{ITS}{Intelligent Transportation System}
	\acro{TL}{Transfer Learning}
	\acro{HD}{High Definition}
	\acro{HSPA}{High Speed Packet Access}
	\acro{LSTM}{Long Short Term Memory}
	\acro{LTE}{Long Term Evolution}
	\acro{ML}{machine Learning}
	\acro{MTL}{Multi-Task Learning}
	\acro{MCS}{Modulation and Coding Scheme}
	\acro{NSA}{Non-Standalone}
	\acro{HVPM}{High voltage Power Monitor}
 \acro{PUE}{Pedestrian UE}
 \acro{PL}{ Platoon Leader}
 \acro{PM}{ Platoon Member}
	\acro{QoS}{Quality of Service}
	\acro{QoE}{Quality of Experience}
	\acro{RB}{Resource Block}
	\acro{RF}{Random Forest}
	\acro{RFL}{Random Forest}
	\acro{RL}{Reinforcement Learning}
	\acro{RRC}{Radio Resource Control}
	\acro{RSSI}{Received Signal Strength Indicator}
	\acro{RSRP}{Reference Signal Received Power}
	\acro{RSRQ}{Reference Signal Received Quality}
	\acro{SCS}{Sub-carrier \ Spacing}
	\acro{SINR}{Signal-to-Interference-Plus-Noise-Ratio}
	\acro{SNR}{Signal-to-Noise-Ratio}
	\acro{SE}{spectral efficiency}
	\acro{UE}{User Equipment}
	\acro{UHD}{Ultra HD}
	\acro{VUE}{Vehicular UE}
	\acro{VoLTE}{Voice over LTE}
	\acro{RNN}{Recurrent Neural Network}
	\acro{WiFi}{Wireless Fidelity}
	\acro{ARIMA}{Auto Regressive Integrated Moving Average}
	\acro{ML}{Machine Learning}
	\acro{NR}{New Radio}
	\acro{Non-IID} {Independent and Identically Distributed}
 \acro{OFDMA}{Orthogonal Frequency Division Multiple Access}
	\acro{TP}{Throughput Prediction}
	\acro{CQI}{Channel Quality Indicator}
	\acro{URLLC}{Ultra Reliable Low Latency Communications}
	\acro{V2X}{Vehicle-to-Everything}
	\acro{C-V2X}{Cellular Vehicle-to-Everything}
	\acro{V2I}{Vehicle-to-Infrastructure}
	\acro{V2V}{Vehicle-to-Vehicle}
    \acro{OFDM}{Orthogonal Frequency Division Multiplexing }
    \acro{BWP}{Bandwidth Part}
    \acro{PUCCH}{Physical Uplink Control Channel}
    \acro{TTL}{Time-to-Live}
    \acro{RA}{resource allocation}
    \acro{BLER}{Block Error Rate}
    \acro{LA}{Link Adaptation}
    \acro{RC}{Resource Chunk}
    \acro{RBs}{Resource Blocks}
    \acro{TTI}{Transmission Time Interval}
    
    \acro{ACC}{Adaptive Cruise Control}
    \acro{CACC}{Cooperative Adaptive Cruise Control}
    \acro{CAV}{Co-operative Automated Vehicle}
    \acro{CAM}{Cooperative Awareness Messages}
    \acro{C-V2X}{Cellular vehicle-to-everything}
    \acro{IoV}{Internet of Vehicles}
    \acro{LTE}{Long Term Evolution}
    \acro{NR}{New Radio}
    \acro{OFDMA}{Orthogonal Frequecy Division Multiple Access}
    \acro{V2V}{vehicle-to-vehicle}
    \acro{V2I}{vehicle-to-infrastructure}
    \acro{V2P}{vehicle-to-pedestrian}
    \acro{eV2X}{enhanced Vehicle-to-Everything}
    \acro{MaxC/I}{Maximum Carrier to Interference Ratio}
    \acro{DRR}{Deficit Round Robin}
    \acro{PF}{Proportional Fair}
    \acro{AoI}{Age of Information}
    \acro{E2E}{End-to-End}
    \acro{IFT}{Information Flow Topologies}
    \acro{PL}{Platoon Leader}
    \acro{PT}{Platoon Tail}
    \acro{PM}{Platoon Member}
    \acro{PLF}{Predecessor-Leader Follower}
    \acro{IP}{Internet Protocol}
    \acro{UE}{User Equipment}
    \acro{FIFO}{First In First Out}
    \acro{MAC}{Media Access Control}
    \acro{UL}{uplink}
    \acro{DL}{downlink}
\end{acronym}

\bibliographystyle{IEEEtran}
\bibliography{references}

\end{document}